\documentclass[%
superscriptaddress,
 amsmath,amssymb,
floatfix,
twocolumn
]{revtex4-2}

\usepackage[utf8]{inputenc}
\usepackage[T1]{fontenc}
\usepackage{etoolbox}

\usepackage[dvipdfm]{graphicx} 
\usepackage{amsmath}
\usepackage{physics}
\usepackage{xcolor}
\usepackage{bm}
\usepackage{mathtools}
\usepackage[ruled,lined]{algorithm2e}
\usepackage{siunitx}
\usepackage{chemformula}
\usepackage{titlesec}
\usepackage[acronym]{glossaries}
\usepackage{enumitem}
\usepackage{grffile}
\usepackage{booktabs} 
\usepackage{listings}
\usepackage{xcolor}
\usepackage{ulem}
\newcommand{\revision}[1]{\textcolor{blue}{#1}}
\usepackage{cancel}
\usepackage{multirow}
\usepackage{makecell}

\definecolor{codegreen}{rgb}{0,0.6,0}
\definecolor{codegray}{rgb}{0.5,0.5,0.5}
\definecolor{codepurple}{rgb}{0.58,0,0.82}
\definecolor{backcolour}{rgb}{0.95,0.95,0.92}

\lstdefinestyle{mystyle}{
    backgroundcolor=\color{backcolour},   
    commentstyle=\color{codegreen},
    keywordstyle=\color{magenta},
    numberstyle=\tiny\color{codegray},
    stringstyle=\color{codepurple},
    basicstyle=\ttfamily\footnotesize,
    breakatwhitespace=false,         
    breaklines=true,                 
    captionpos=b,                    
    keepspaces=true,                 
    numbers=left,                    
    numbersep=5pt,                  
    showspaces=false,                
    showstringspaces=false,
    showtabs=false,                  
    tabsize=2
}

\newcommand{\vect}[1]{\mathbf{#1}}               

\titleformat*{\paragraph}{\normalfont\bfseries}

\begin{document}

\title{Cross Learning between Electronic Structure Theories for Unifying Molecular, Surface, and Inorganic Crystal Foundation Force Fields}

\author{Ilyes Batatia}\thanks{These authors contributed equally.}
\affiliation{Engineering Laboratory, University of Cambridge, Trumpington St, Cambridge, UK}

\author{Chen Lin}\thanks{These authors contributed equally.}
\affiliation{University of Oxford, UK}
\author{Joseph Hart}
\affiliation{Engineering Laboratory, University of Cambridge, Trumpington St, Cambridge, UK}
\affiliation{Cavendish Laboratory, University of Cambridge, J. J. Thomson Avenue, Cambridge, UK}
\author{Elliott Kasoar}
\affiliation{Scientific Computing Department, Science and Technology Facilities Council, Daresbury Laboratory, Keckwick Lane, Daresbury WA4 4AD, UK}
\affiliation{Engineering Laboratory, University of Cambridge, Trumpington St, Cambridge, UK}
\author{Alin M. Elena}
\affiliation{Scientific Computing Department, Science and Technology Facilities Council, Daresbury Laboratory, Keckwick Lane, Daresbury WA4 4AD, UK}
\author{Sam Walton Norwood}
\affiliation{Mirror Physics, 31 Hudson Yards, Floor 11, New York, NY 10001}
\author{Thomas Wolf}
\affiliation{Hugging Face, 20 Jay Street  Suite 620, Brooklyn, NY 11201}
\author{G\'abor Cs\'anyi}
\affiliation{Engineering Laboratory, University of Cambridge, Trumpington St, Cambridge, UK}

\date{\today}

\begin{abstract}
Creating a single unified interatomic potential capable of attaining \textit{ab initio} accuracy across all chemistry remains a long-standing challenge in computational chemistry and materials science. This work introduces a training protocol for foundation machine-learning interatomic potentials (MLIPs) that bridge molecular, surface, and materials chemistry through cross-domain learning. First, we introduce enhancements to the MACE architecture that improve its performance on chemically diverse databases by increasing weight sharing across chemical elements and introducing non-linear factors into the tensor decomposition of the product basis. Second, we develop a multi-head replay post-training methodology that enables efficient knowledge transfer across diverse chemical domains. By fine-tuning on datasets at different levels of electronic structure theory—including inorganic crystals, molecular systems, surface chemistry, and reactive organic chemistry—we demonstrate that a single unified model achieves state-of-the-art performance across several chemical domains. Comprehensive benchmarking reveals superior cross-domain transferability compared with existing specialised and multi-task models, with notable improvements in molecular and surface properties while maintaining state-of-the-art performance in materials-property prediction.
\end{abstract}

\maketitle
\def\thefootnote{*}\footnotetext{These authors contributed equally.}

\section{Introduction}
\label{sec:intro}

The development of accurate and transferable machine learning interatomic potentials (MLIPs) remains one of the most challenging problems in computational chemistry and materials science~\cite{behler2007,gap,THOMPSON2015316,schnet,Smith2017,deringer_machine_2019,drautz2019,lilienfeld2020,batzner2022,batatia_mace_2023,ko2023recent}. Traditional approaches have created a fragmented landscape where distinct models are required for molecular systems~\cite{Smith2017, Devereux2020, Anstine2025, Kovcs2025}, surface chemistry~\cite{Chanussot2021, tran2023open, sahoo2025opencatalyst2025oc25}, and bulk materials~\cite{batatia2024foundationmodelatomisticmaterials,Chen2022-qb,merchant_scaling_2023}, creating substantial barriers when studying phenomena that naturally span multiple chemical domains, such as heterogeneous catalysis, crystal growth, or interfacial processes.
Recent advances in foundation MLIPs have demonstrated remarkable capabilities through large pre-training on diverse datasets~\cite{batatia2024foundationmodelatomisticmaterials,Chen2022-qb,merchant_scaling_2023,Deng2023-qn,yang2024mattersim, mazitov2025petmadlightweightuniversalinteratomic, Zhang2024}. However, most existing foundation models suffer from limited chemical scope, often excelling in one domain while showing inadequate performance in others. This limitation stems from both dataset constraints and architectural choices that favour domain-specific optimisation over cross-domain transferability.

The challenge of unifying multiple chemical domains within a single MLIP framework involves several key considerations: (i) training on datasets with different levels of electronic structure theory, (ii) efficient knowledge transfer between chemical domains without catastrophic forgetting (if training involves multiple stages), and (iii) maintaining computational efficiency while expanding chemical coverage. One strategy, used recently   by UMA~\cite{wood2025umafamilyuniversalmodels},  DPA-3~\cite{zhang2025graphneuralnetworkera} and SevenNet~\cite{Kim2024}, is to make the model output explicitly depend on the task by embedding the task as part of the input along with the atomic coordinates. This makes most model layers task-dependent, and allows for significant flexibility while benefiting from  some cross-learning. An alternative is meta-learning, where during a pre-training stage multiple tasks are sequentially attempted, resulting in a model that is easy to specialise to a specific task by performing fine-tuning training~\cite{Allen2024}. The downside of both these approaches is that at inference time the task needs to be specified by the user; there is no single model applicable to all tasks out of the box. DPA-2~\cite{Zhang2024} and JMP~\cite{shoghi2024from} used pre-training with multiple readout heads (where only the last computational layer is specific to each task) and downstream fine-tuning for a given specific task. In this work, we create a single foundation MLIP using multiple diverse datasets, also using a multi-head approach, and evaluate the performance of its ``main'' head (in this case corresponding to DFT with the PBE functional) on {\em all} tasks. The ultimate goal is to have a single continuous potential energy function applicable to all chemical contexts. We therefore design a training protocol that enhances cross-learning and knowledge sharing from all heads to the main head.

Our first contribution is to introduce a set of changes to the state-of-the-art MACE architecture that improves its performance for large databases containing a large number of chemical species. These changes, explicitly highlighted below, include the use of more weight-sharing between the different species to enable the model to learn more powerful compressions of the chemical domains. We also introduce non-linear factors in the tensor decomposition that are demonstrated to provide better accuracy than purely polynomial features.

Second, we introduce a multi-head cross-learning fine-tuning protocol that aims to produce a single model bridging materials, molecular, and surface chemistry through two key innovations: (1) a multi-head architecture that enables simultaneous learning across potentially inconsistent levels of electronic structure theory while maintaining shared chemical and geometrical representations; (2) pre-training followed by a replay fine-tuning methodology that facilitates knowledge transfer across domains while preventing catastrophic forgetting from the base model. 
We perform comprehensive validation across molecular, surface, and materials benchmarks, establishing new standards for evaluating unified foundation force field accuracy.

Our results show that cross-domain learning attains competitive in-domain accuracy and yields measurable cross-domain generalisation via knowledge transfer, improving performance on molecular, surface, and crystal benchmarks without degrading materials accuracy. This work establishes a clear path toward MLIPs that can seamlessly handle physical and chemical phenomena across all areas of chemistry and materials science.

\section{Methods}
\label{sec:methodology}

\subsection{Theoretical Foundation}

Our approach builds upon the MACE architecture \cite{batatia_mace_2023}, which employs many-body equivariant message passing to build fast and accurate machine learning interatomic potentials. 
MACE parameterises a mapping from atomic coordinates and atomic numbers (element indices) to the potential energy by decomposing it into atomic energies.
Each atomic energy term is a function of invariant descriptors that embed the chemical and geometrical many-body information of the neighbourhood of the atom.
The total energy $E$ of a system is expressed as:
\begin{equation}
E = \sum_i E_{i}(\{\vect{r}_{ij}, z_j\}_{j \in \mathcal{N}_i})
\end{equation}
where $E_{i}$ represents the atomic energy contribution of atom $i$, $\mathcal{N}_i$ denotes the set of its neighbours within a cutoff radius $r_\text{cut}$, $\vect{r}_{ij}$ are relative vectors from atom $i$ to its neighbour $j$, and $z_j$ are atomic numbers.

The multi-head architecture enables simultaneous learning across multiple potentially inconsistent levels of electronic structure theory by employing distinct shallow readout functions that map shared latent feature representations to each desired theoretical framework:
\begin{equation}
E_{i}^{(\text{head})} = \sum_s \mathcal{R}^{(\text{head}, s)}(\vect{h}_i^{(s)}) + E_{0, z_{i}}^{(\text{head})}
\label{eq:multihead_energy}
\end{equation}
where $(\text{head})$ enumerates readout  heads corresponding to different levels of theory, $\vect{h}_i^{(s)}$ represents the node features at layer $s$, and $E_0^{(\text{head})}$ are head-specific atomic reference energies. For all the fine-tuning heads, we use the method outlined in Section~\ref{sec:e0-est} to estimate their $E_{0}$, and for the superior cross-domain transferability head we use DFT computed atomic reference energies. This multi-head approach parallels recent developments in machine learning potentials, including the DPA-2 model~\cite{Zhang2024}. In our case, for each head, we use a simple linear readout layer for the first layer and a single-hidden-layer fully-connected feedforward network (multilayer perceptron, MLP) for the second layer.

\subsection{MACE with non-linear tensor decomposition}
\label{sec:non-lin-block}

Equations \eqref{eq:element_embedding}-\eqref{eq:update} represent the complete functional form of MACE as they were introduced in Ref.~\cite{batatia_mace_2023, Kovcs2023} together with the blue parts corresponding to the modifications we introduce in this paper. We arrived at this particular set of changes by performing a grid-search over proposed changes using large-dataset training runs to maximise the accuracy of the model with the fewest changes. For ease of reference, we keep the formulas together and provide a detailed explanation of them below. Note that in order to help readability by limiting the number of symbols and symbol modifiers, we reuse the symbol $W$ to refer to the free parameters of the model; each occurrence below corresponds to independent weights with appropriate dimensions and indices. There are additional free parameters in the MLPs which are not explicitly shown.

The finite neighbourhood of each atom creates a graph topology on the atomic structures. Each node $i$ carries a feature vector $\bm h_i$, expanded in a spherical-harmonic basis so that components are indexed by $(l,m)$; we write $\bm h_i^{(s)}$ for the features after iteration $s$ (the $s$-th message-passing layer) and denote the total number of layers by $S$.

We start by initialising the node features ${\bm h}^{(0)}_{i}$ as a learnable embedding of the chemical elements with atomic number $z_i$ into $K_{\text{node}}$ learnable channels indexed by $k$, cf. Eq.~\eqref{eq:element_embedding}. This kind of mapping has been used extensively for graph neural networks~\cite{schutt2017schnet, schutt2021PAINN, batzner20223NequIP, gasteiger2020DimeNet++} and elsewhere~\cite{willatt2018feature, gubaev2019accelerating} and has been shown to lead to some transferability between molecules with different elements~\cite{batatia2022designSpace}. The zeros for the $lm$ indices correspond to these initial features being scalars, i.e. rotationally invariant. The higher-order elements of ${\bm h}^{(0)}_{i}$ with nonzero $lm$ indices are initialised to zero.
At the beginning of each subsequent iteration, the node features (both scalars and higher order) are linearly mixed together resulting in ${\bm{\bar h}}_{j}$, cf. Eq.~\eqref{eq:linear_first}.

Next, we combine the features of each of the neighbouring atoms $j$ with the interatomic displacement vectors pointing to them from the central atom $i$ (corresponding to the $i$--$j$ edge in the graph) expressed using radial and spherical harmonic basis. This is analogous to the construction of the one-particle basis of neighbour density representations, such as SOAP~\cite{bartok2013SOAP} and ACE~\cite{drautz2019,batatia2022designSpace}, and we construct it in a similar way to Cormorant~\cite{anderson2019cormorant} and NequIP~\cite{batzner20223NequIP}. The relationships between these different approaches are discussed in Ref.~\cite{batatia2022designSpace}. We construct the radial basis set using the first spherical Bessel function $j^n_0$ for different wavenumbers, $n$, up to some small maximum (typically 8), in Eq.~\eqref{eq:rad_feats} as proposed in Ref.~\cite{gasteiger2020DimeNet++}. 
For each channel $k$, the radial information is then passed through a separate MLP, Eq.~\eqref{eq:radial_MLP}, whose inputs are the Bessel functions with different frequencies, and which has many outputs, indexed by $(\eta_1,l_1,l_2,l_3)$. In a modification to the original MACE architecture, we compute separate embeddings of the source and target elements following~\cite{wood2025umafamilyuniversalmodels}, and concatenate them with the Bessel features before passing them into the radial MLP.
The next modification compared to the original MACE design is that we apply the radial cutoff function $f_\text{cut}$ outside the MLP, and not directly to the Bessel function as in the original MACE. This forces a smoother decay near the cutoff.
When combining positional information (itself an equivariant that transforms under rotation like a vector) with equivariant node features, we use the spherical tensor product formalism of angular momentum addition~\cite{wigner2012group}. All possible combinations of equivariants are constructed using the appropriate Clebsch-Gordan coefficients, Eq.~\eqref{eq:phi-basis-t}. 

The one-particle basis $\phi$ is summed over the atoms in the neighbourhood in Eq.~\eqref{eq:atomic-basis-t}. This is where permutation invariance of the MACE descriptors over the atoms in the neighbourhood is achieved - note that the identity of chemical elements has already been embedded, and hence this sum is over all atoms, regardless of their atomic number. A linear mixing of $k$ channels with learnable weights yields the initial atomic basis, $\tilde A_i$. As the tensor product operation in Eq.~\eqref{eq:phi-basis-t}, which happens on the edges, is the computational bottleneck of MACE, we let the freedom to use fewer channels for this operation and use $K_{\text{edge}}^{(s)} \leq K$  instead. The dependence $s$ is used to specify more edge channels for the inexpensive first layer $s=0$ compared to other layers.

To ensure internal normalization of the features and smooth extrapolation to systems with different densities, we divide the atomic basis in each layer by a learnable quantity called density normalization $n_{i}$. We extend our density normalization~\cite{batatia2024foundationmodelatomisticmaterials} to be a (learnable) weighted sum of the density term and a constant term. This weighted sum acts like a mixture of a body-ordered feature (with the constant term) and a mean-field feature with the density term.

\begingroup\makeatletter\def\f@size{8.5}\check@mathfonts
\def\maketag@@@#1{\hbox{\m@th\large\normalfont#1}}%
\begin{align}
\label{eq:element_embedding}
h_{i,k00}^{(0)} &= \sum_z W_{kz} \delta_{zz_{i}}\\
\vbox to 20pt{}
\label{eq:linear_first}
    \bar{h}^{(s)}_{i,kl_2m_2} &= \sum^{K_{\text{node}}}_{\tilde{k}} W_{k\tilde{k}l_2}^{(s)} h^{(s)}_{i,\tilde{k}l_2m_2} \\
\label{eq:rad_feats}
j^{n}_{0} (r_{ij}) &=  \sqrt{\frac{2}{r_{\text{cut}}}} \frac{\sin{\left(n\pi\frac{r_{ij}}{r_{\text{cut}}} \right)}}{r_{ij}} \revision{\xcancel{f_{\text{cut}}(r_{ij})}} \\
\label{eq:source_embedding}
\revision{e_{i, k}^{(s), \text{src}}} &\revision{= \sum_z W_{kz} \delta_{zz_{i}}, \quad e_{i, k}^{(s), \text{trg}} = \sum_z W_{kz} \delta_{zz_{i}}} \\
\vbox to 20pt{}
\label{eq:radial_MLP}
    R_{k \eta_{1} l_{1}l_{2} l_{3}}^{(s)}(r_{ij}) &=   {\rm MLP}\left( \left\{ {j_0^n} (r_{ij})\right\}_{n}, \revision{e_{i}^{(s), \text{src}}, e_{j}^{(s), \text{trg}}}\right) \revision{f_{\text{cut}}(r_{ij})} \\
\vbox to 20pt{}
  \label{eq:phi-basis-t}
  \phi_{ij,k \eta_{1} l_{3}m_{3}}^{(s)} &= 
    \sum_{l_1l_2m_1m_2} C_{\eta_1,l_1m_1l_2m_2}^{l_3m_3}
      R_{k \eta_{1} l_{1}l_{2} l_{3}}^{(s)}(r_{ij}) \,\,\times \notag\\
      & \qquad\qquad \times Y^{m_{1}}_{l_{1}} (\boldsymbol{\hat{r}}_{ij}) \bar{h}^{(s)}_{j,kl_2m_2} \\ 
\vbox to 20pt{}
\label{eq:normalization}
\revision{n_{i}^{(s)}}  \revision{=\sum_{j \in \mathcal{N}(i)}}& \revision{\tanh \left(  {\rm MLP}\left( \left\{ {j_0^n} (r_{ij})\right\}_{n}, \revision{e_{i}^{(s), \text{src}}, e_{j}^{(s), \text{trg}}}\right)^2 \right)f_{\text{cut}}(r_{ij})} \\
\vbox to 20pt{}
\label{eq:atomic-basis-t}
     A_{i,kl_{3}m_{3}}^{(s)} & \revision{=\frac{1}{\alpha^{(s)} + \beta^{(s)} n_{i}^{(s)}}}\sum_{ \eta_{1}}\sum_{\tilde{k}}^{\revision{K_{\text{edge}}^{(s)}}} W_{k \tilde{k} \eta_{1}l_{3}}^{(s)}
    \sum_{j \in \mathcal{N}(i)}  \phi_{ij,\tilde{k} \eta_{1} l_{3}m_{3}}^{(s)} \\
\vbox to 20pt{}
\label{eq:gate-eq-res}
    \revision{\tilde A_{i,kl_{3}m_{3}}^{(s)}  } & \revision{ =A_{i,kl_{3}m_{3}}^{(s)} + \sum^{K_{\text{node}}}_{\tilde{k}} W_{k\tilde{k}l_3}^{(s)} \bar{h}^{(s)}_{i,\tilde{k}l_3m_3}} \\
\vbox to 20pt{}
\label{eq:atomic-basis-scalars}
     A_{i,kl_{3}}^{(s), \text{scalars}} &= \revision{\frac{1}{\alpha^{(s)} + \beta^{(s)}n_{i}^{(s)}}}\sum_{ \eta_{1}}\sum_{\tilde{k}}^{\revision{K_{\text{edge}}^{(s)}}} W_{k \tilde{k} \eta_{1}l_{3}}^{(s)}
    \sum_{j \in \mathcal{N}(i)}  \phi_{ij,\tilde{k} \eta_{1} 00}^{(s)} \\
\vbox to 20pt{}
\label{eq:gate-scalars-res}
    \revision{\Omega_{i,kl_{3}}^{(s)}  } & \revision{ =A_{i,kl_{3}}^{(s), \text{scalars}} + \sum^{K_{\text{node}}}_{\tilde{k}} W_{k\tilde{k}l_{3}}^{(s)} \bar{h}^{(s)}_{i,\tilde{k}00}} \\
\vbox to 20pt{}
\label{eq:gate-non-lin}
\revision{g(x_{i,k00}, y_{i,klm})} &
\revision{=\begin{cases}
\revision{x_{i,k00}\operatorname{\sigma}\!\left(x_{i,k00}\right)y_{i,k00},} & \revision{l = 0,}\\
\revision{\sigma\!\left(x_{i,k00}\right)y_{i,klm}}, & \revision{l > 0},
\end{cases}} \\
\vbox to 20pt{}
\label{eq:atomic-basis-non-lin}
    \revision{A_{i,kl_{3}m_{3}}^{(s), \text{gated}}} &\revision{= \sum^{K_{\text{node}}}_{\tilde{k}} W_{k \tilde{k}l_{3}}^{(s)}g( \Omega_{i,kl_{3}}, \tilde A_{i,kl_{3}m_{3}}^{(s)})} \\
\vbox to 20pt{}
\label{eq:product-basis}
{\bm A}^{(s),\nu}_{i,k\bm l \bm m} &= \prod_{\xi = 1}^{\nu} A_{i,k l_\xi  m_\xi}^{(s), \revision{\text{gated}}}\\
\vbox to 20pt{}
\label{eq:symmbasis_L1}
  {\bm B}^{(s),\nu}_{i,\eta_{\nu} k LM}
  &= \sum_{{\bm l}{\bm m}} \mathcal{C}^{LM}_{\eta_{\nu} \bm l \bm m} {\bm A}^{(s),\nu}_{i,k\bm l \bm m} \\
 \vbox to 20pt{}
 \label{eq:message}
  m_{i,k LM}^{(s)} &=  \sum_{\nu}\sum_{\eta_{\nu}} W_{\revision{\xcancel{z_{i}}} \eta_{\nu} k L}^{(s),\nu} {\bm B}^{(s),\nu}_{i,\eta_{\nu} k LM}\\
 \vbox to 20pt{}
 \label{eq:update}
  h^{(s+1)}_{i,k LM}
    &= \sum^{K_{\text{node}}}_{\tilde{k}} W_{k L,\tilde{k}}^{(s)} m_{i,\tilde{k}LM}^{(s)}
  + \sum^{K_{\text{node}}}_{\tilde{k}} W_{k\revision{\xcancel{z_{i}}}L,\tilde{k}}^{(s)}
  h^{(s)}_{i,\tilde{k}LM} 
\end{align}\endgroup

We then construct an updated atomic basis, $A^\text{gated}_i$ in Eq.~\eqref{eq:gate-eq-res}, using a learnable residual connection from the initial node features,  that contains both information about the neighbourhood of the atom and also information about the atom $i$ itself. This new modification enables the model to construct a polynomial that contains not only factors from the neighbours of $i$, but also factors that depend on the features of $i$ directly, allowing for richer many-body features.

We also construct a set of scalar features $\Omega$ on each atom, from a learnable combination of extra scalars computed from the atomic basis $A_{i}^{(s), \text{scalars}}$ in Eq.~\eqref{eq:atomic-basis-scalars} and the node features as a residual connection in Eq.~\eqref{eq:gate-scalars-res}. The scalar features are used to compute a gated non-linearity~\cite{geiger2022e3nneuclideanneuralnetworks} in equations~\eqref{eq:gate-non-lin} and \eqref{eq:atomic-basis-non-lin}. We found that a sigmoid ($\sigma$) gate for the equivariant channels and a SiLU for the invariant channels works best.

We use these gated atomic basis to construct many-body messages in the same way as in our original MACE design using tensor-decomposed symmetric contractions. For more details on these operations, see the original MACE paper~\cite{batatia_mace_2023}. Additionally, for more explanation on the role of the tensor decomposition, refer to the TRACE paper~\cite{PhysRevLett.131.028001}. One important modification compared to the original tensor decomposition is that, due to the non-linearity in Eq.~\ref{eq:atomic-basis-non-lin}, the model can learn non-linear rank-1 factors in the tensor decomposition. We also use element-agnostic weights for the message construction (Eq.~\eqref{eq:message}) and the update (Eq.~\eqref{eq:update}). The rest of the architecture is identical to that of our previous MACE models, with the computation of the product basis in Eq.~\ref{eq:product-basis}, the symmetrization with the generalised Clebsch–Gordan coefficients in Eq.~\ref{eq:symmbasis_L1}, the message construction in Eq.~\ref{eq:message} and the update in Eq.~\ref{eq:update}. After S layers of message passing (usually we use S=2), the node features are used to predict site energies per head in Eq.~\ref{eq:multihead_energy}.

\section{Multi-Head Replay Post-Training}
\label{sec:replay}

\begin{figure*}[!t]
\centering
\includegraphics[width=0.9\textwidth]{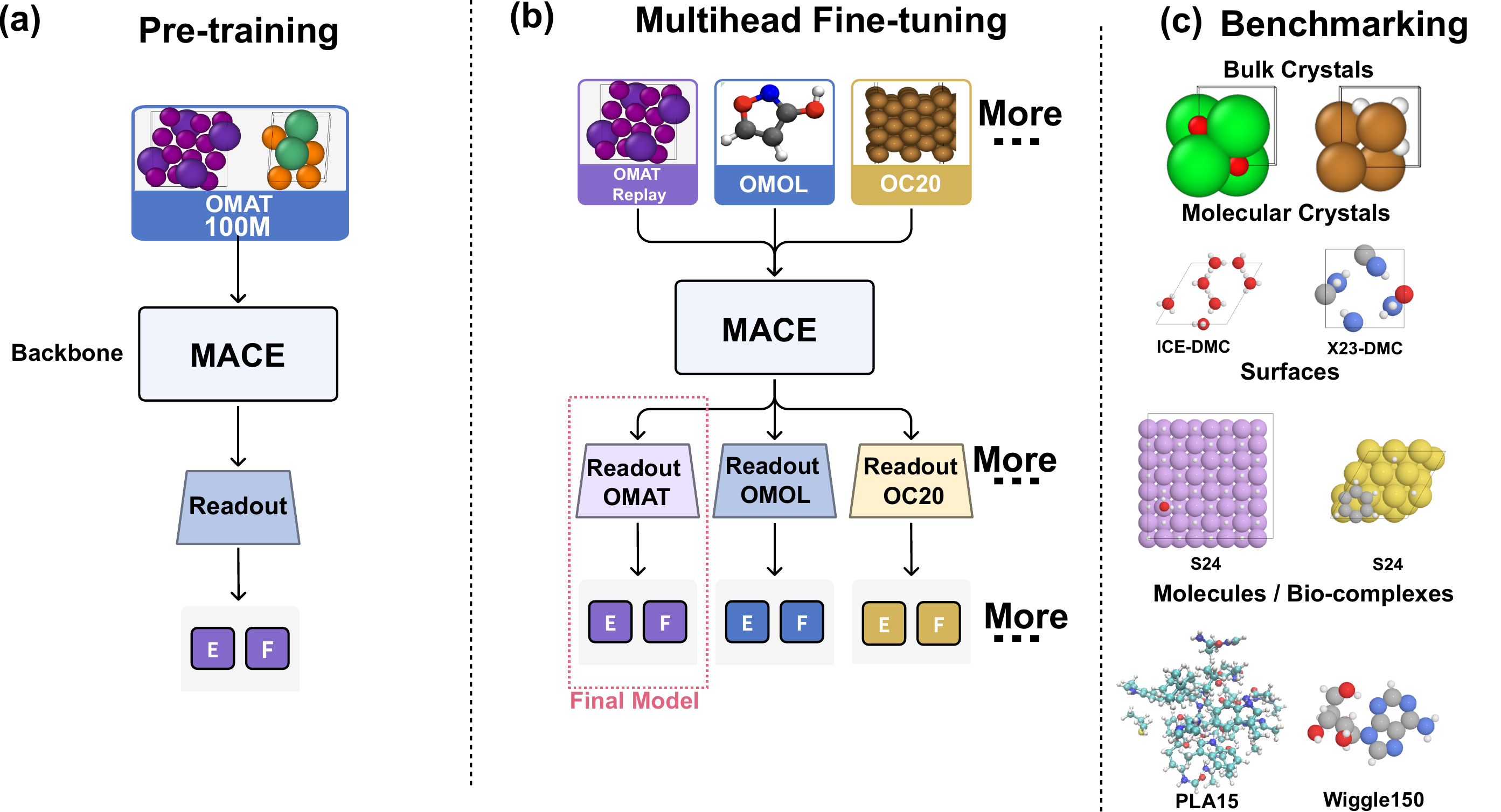}
\caption{\label{fig:methodology_workflow}\textbf{Workflow for cross-domain machine learning interatomic potential development.} (a) Stage 1 establishes foundation chemical knowledge through pre-training on large-scale inorganic materials data (OMAT-24). (b) Stage 2 implements multi-head fine-tuning with strategic replay across diverse chemical domains, enabling knowledge transfer while preventing catastrophic forgetting. (c) The resulting unified model is benchmarked across molecular, materials, and surface chemistry tests and achieves state-of-the-art performance.}
\end{figure*}

Figure~\ref{fig:methodology_workflow} provides an overview of our cross-learning framework, illustrating the progression from foundation pre-training through Multi-Head Replay Post-Training.
Our multi-head replay post-training strategy~\cite{batatia2024foundationmodelatomisticmaterials} facilitates efficient knowledge transfer from a broad foundation dataset to multiple specialised datasets while preventing catastrophic forgetting through strategic replay sampling. The workflow comprises two distinct stages:

\begin{itemize}
    \item \textbf{Stage 1: Pre-training at base level theory.} We first train a unified backbone model on a large, diverse dataset of inorganic crystals, producing shared feature representations that capture fundamental chemical and geometrical patterns.
    \item \textbf{Stage 2: Multi-head Fine-tuning with Replay.} We simultaneously fine-tune multiple shallow readout heads, starting from the pre-trained weights on domain-specific datasets. We do not freeze any weights, i.e. we keep fine-tuning the backbone weights as well. Each head targets a different chemical domain (molecular systems, surfaces, inorganic materials) or specific level of theory. To mitigate catastrophic forgetting, we construct a replay buffer by sampling representative configurations from the original pre-training data. During fine-tuning, each minibatch combines randomly new domain-specific samples with replay samples, ensuring retention of foundational chemical knowledge.
\end{itemize}

\section{Datasets}
\label{sec:datasets}

We employ a combination of large-scale foundation data and targeted fine-tuning datasets to comprehensively cover inorganic crystals, surfaces, and molecular chemistry.

\subsection{Pre-training Dataset}

\textbf{OMAT}~\cite{barrosoluque2024openmaterials2024omat24}: A large inorganic crystal dataset containing 100 million configurations spanning 89 elements. All calculations are performed at the PBE and PBE+$U$ level of theory, providing broad coverage of diverse inorganic crystals and serving as our pre-training dataset. 

\subsection{Fine-tuning Datasets}

\begin{itemize}
    \item \textbf{OMAT Replay} (10\% subset): We randomly selected 10 million configurations from OMAT as a replay buffer to prevent catastrophic forgetting of foundational inorganic structures during fine-tuning. We will keep this head as our PBE head for the model.
    \item \textbf{RGD1}~\cite{Zhao2023}: Contains 300K configurations of small organic reaction intermediates and transition states computed at the B3LYP/6-31G* level, enabling accurate modelling of organic reaction pathways.
    \item \textbf{MPTraj}: This dataset comprises 1.5 million configurations from the Materials Project (MP) \cite{Horton2025} including static calculations and structural optimisation trajectories. The dataset emphasises dynamic lattice distortions in small periodic unit cells (90\% under 70 atoms) describing inorganic crystals with some molecular components. DFT calculations employ the PBE exchange-correlation functional with Hubbard $U$ terms for selected transition metal oxides~\cite{MP_calc_details}. This dataset was originally compiled for CHGNet~\cite{Deng2023-qn}.
    \item \textbf{SPICE-1}~\cite{Eastman2023}: Encompasses $\sim$500K geometries of small to medium-sized organic molecules calculated at the $\omega$B97M-D3(BJ)/def2-TZVP level, providing comprehensive coverage of conformational landscapes and intramolecular interactions.
    \item \textbf{OC20}~\cite{Chanussot2021}: We randomly subsample 2 million metal surface slabs and adsorbate complexes computed at the PBE level, specifically targeting catalytic surface processes and gas-surface interactions.
    \item \textbf{OMOL-1\%}~\cite{levine2025openmolecules2025omol25}: Subsampled 1.2 million neutral, closed-shell and diverse organic, organometallic, and transition-metal configurations, ensuring comprehensive coverage of coordination chemistries, calculated using hybrid DFT with the $\omega$B97M-VV10 functional.
    \item 
    \textbf{MATPES R2SCAN} \cite{kaplan2025foundationalpotentialenergysurface}: Comprises 400K inorganic crystal snapshots sampled via molecular dynamics using machine learning force fields at the r$^2$SCAN level (without Hubbard $U$ corrections).
\end{itemize}

\section{Model Descriptions}

The hyperparameters for all MACE models can be found in the supplementary Table~\ref{tab:mace_hparams}. We benchmark several models to evaluate the effectiveness of our multi-head replay approach:

\begin{itemize}
    \item \textbf{mace-omat-1}: 
    Initial pre-training MACE model on the full OMAT dataset. It uses the new block proposed in~\ref{sec:non-lin-block}.
    \item \textbf{mace-mh-1}: Our proposed MACE model, fine-tuned with multi-head replay fine-tuning on the mace-omat-1 backbone. All of the results in the main text use the single ``OMAT head'' of the model referred to as \textbf{mace-mh-1-omat}. We also benchmark the ``OMOL'' head fine-tuned on the OMOL dataset referred to as \textbf{mace-mh-1-omol-1\%}. The $\textbf{1\%}$ reflects that it was trained on just 1\% of the full OMOL dataset. We also present results for R2SCAN heads of our models in Table~\ref{tab:mh_r2scan_appendix}. For the hyper-parameters of the MACE models, please see {Table~\ref{tab:mace_hparams}}.
    \item \textbf{mace-omat-0}: MACE model trained on the full OMAT dataset, using the original blocks of MACE, with slightly smaller model size (corresponding to a medium sized model, see the supplementary Table~\ref{tab:mace_hparams}).
    \item \textbf{mace-omol}: MACE model trained on the full 100M OMOL dataset, with total charge and total spin global embedding (denoted in the text as \textbf{mace-omol-100\%}). As this model is only trained on molecular systems, we only benchmark it when appropriate.
    \item \textbf{uma}~\cite{wood2025umafamilyuniversalmodels}: eSEN~\cite{fu2025learningsmoothexpressiveinteratomic} model trained on 100M inorganic crystals of OMAT, 230M surfaces/small molecules of OC20/OC22, 100M molecular configurations of OMOL, 25M molecular crystals configurations of OMC and 29M metal organic frameworks of ODAC. The dataset types are embedded as one-hot encoding in the model as a global input. We benchmark the OMAT and OMOL variant throughout this paper, with both the S-1.1 and the larger M-1.1 variant referred to as \textbf{uma-s-1p1-omat/omol-100\%} and \textbf{uma-m-1p1-omat/omol-100\%} respectively in the text. We use omol-100\% to highlight the fact that it is trained on the full dataset.
    \item \textbf{mace-mp-0a}~\cite{batatia2024foundationmodelatomisticmaterials}: MACE trained on the MPTraj dataset, representing a strong baseline for inorganic materials modelling.
    \item \textbf{mattersim-5M}~\cite{yang2024mattersim}: A recent foundation model trained on diverse chemical systems, providing state-of-the-art comparison.
    \item \textbf{orb-v3}~\cite{rhodes2025orbv3atomisticsimulationscale}: Orb model trained on the OMAT AIMD subset; we use the most accurate orb-v3-conservative-inf-omat in all the tests, referred to as \textbf{orb-v3-consv-inf-omat}.
\end{itemize}

\section{Benchmark Overview}
\label{sec:global_overview}

\begin{figure*}[!t]
\centering
\includegraphics[width=\textwidth]{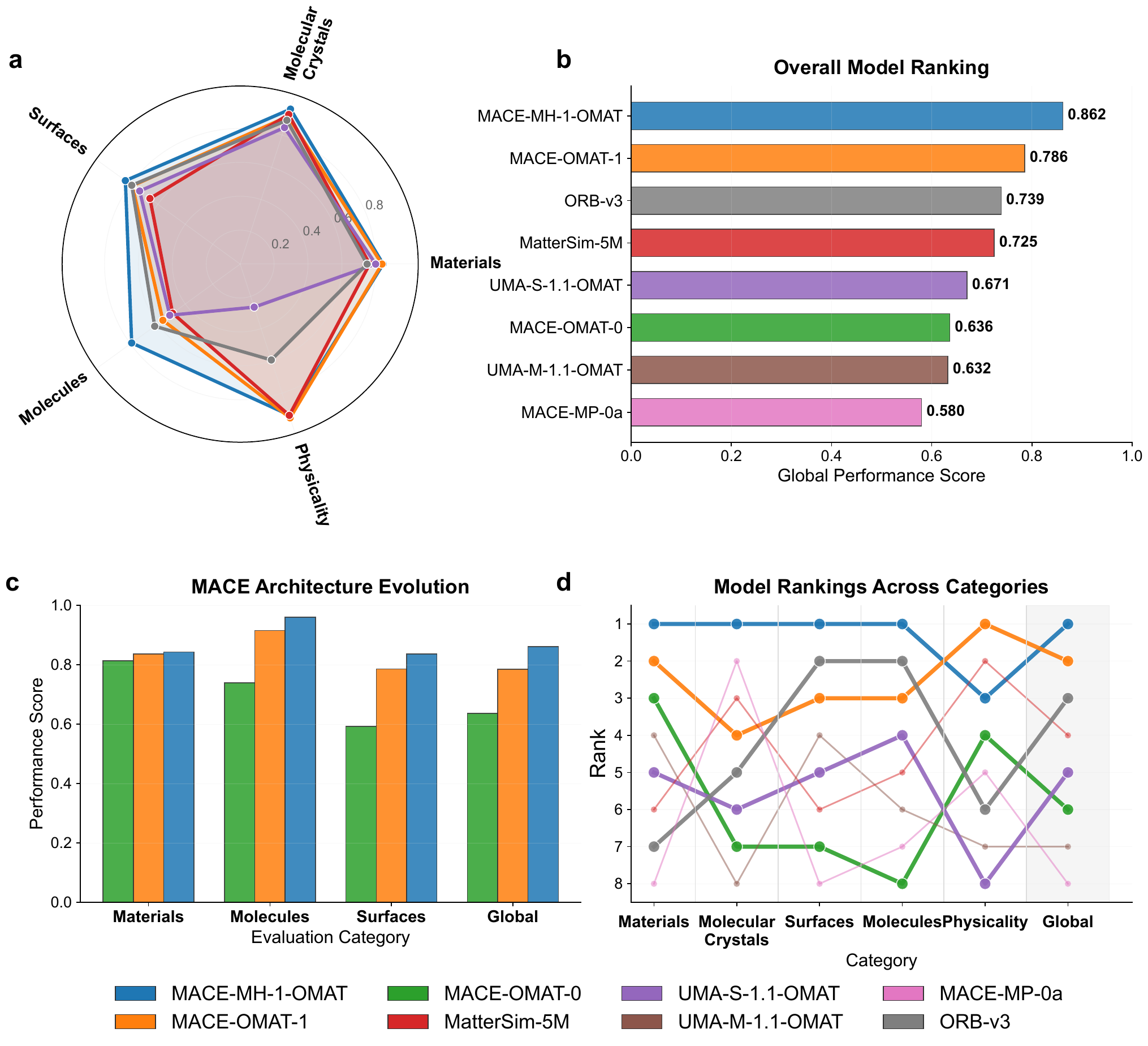}
\caption{\textbf{Cross-domain performance summary of foundation interatomic potentials.} (a) Polar chart of \emph{domain scores} for five evaluation groups—Materials, Molecular Crystals, Surfaces, Molecules, and Physicality—where values are normalised to $[0,1]$ (1 = best; higher is better) and computed as per-metric means within each group. (b) Overall ranking by a \emph{global score}, defined as a weighted sum of domain scores (Materials 0.25, Molecules 0.25, Surfaces 0.20, Molecular Crystals 0.20, Physicality 0.10). (c) Ablation within the MACE family comparing the baseline linear block (mace-omat-0), the non-linear block (mace-omat-1), and the multi-head model (mace-mh-1-omat), shown across Materials, Molecules, Surfaces, and the resulting global score. (d) Per-category model ranks (smaller is better; axis inverted so the top is rank 1) illustrating consistency across domains; the shaded region marks the global rank. The scoring procedure is described in Sec.~\ref{sec:global_overview} and normalisation bounds and metric definitions are given in Table~\ref{tab:normalization_bounds}.}
\label{fig:aggregated-metrics}
\end{figure*}

Figure~\ref{fig:aggregated-metrics} summarises cross-domain performance using five normalised domain scores—\textbf{Materials}, \textbf{Molecular Crystals}, \textbf{Surfaces}, \textbf{Molecules}, and \textbf{Physicality}—and a global score defined as a weighted sum (Materials 0.25, Molecules 0.25, Surfaces 0.20, Molecular Crystals 0.20, Physicality 0.10). We evaluate model physicality through several benchmarks assessing size extensivity, additivity, and
smoothness of dimer interactions, see the Section~\ref{sec:physicality_benchmark} for details. The breakdown of the weighting of each benchmark towards the scores can be found in Table~\ref{tab:normalization_bounds}. We acknowledge that aggregating metrics is a difficult task, and our proposed weighting is necessarily subjective. To normalise the scores, we defined two sets of bounds, one for a model that would be deemed inaccurate and one bound for the accuracy at which the benchmark would be saturated, either because it would reach the intrinsic accuracy of the DFT in the case of comparison with wavefunction methods, or because it is a mathematical upper bound. We believe that going forward, these weights need to be improved by a joint discussion of the community.  Panel~\textbf{(a)} shows a radar plot comparison of domain scores for the leading models. The multi-head model, \textbf{mace-mh-1-omat}, attains the most balanced profile, with strong Molecules and Molecular Crystals while remaining competitive on Materials, Surfaces, and Physicality. Panel~\textbf{(b)} ranks models by the global score and highlights the same model as the top overall performer, indicating that gains in molecular chemistry do not come at the expense of bulk materials or surface physics. Panel~\textbf{(c)} isolates the architecture trajectory within the MACE family—\textbf{mace-omat-0} (linear block) \(\rightarrow\) \textbf{mace-omat-1} (non-linear block) \(\rightarrow\) \textbf{mace-mh-1-omat}—and shows monotonic improvements in the Molecules and Surfaces categories together with stable Materials performance, which together drive the increase in the global score.  The improvements from \textbf{mace-omat-0} to \textbf{mace-omat-1} are due to two factors: a choice of a slightly larger model size, going from $L=1$ messages to $L=2$, and to the changes to the mace architecture outlined in the previous section.
Panel~\textbf{(d)} plots per-category ranks (lower is better; axis inverted), revealing that the best models are not specialists: they maintain good ranks across all five categories, with especially consistent behaviour for \textbf{mace-mh-1-omat}. 

In the following sections, we give a detailed performance breakdown of the different tested models on each benchmark, with materials benchmarks in section~\ref{sec:materials_benchmarks}, molecular crystals in section~\ref{sec:molecular_crystals}, surfaces in section~\ref{sec:surface_benchmarks}, molecular systems in section~\ref{sec:molecular_benchmarks}, physicality benchmark in section~\ref{sec:physicality_benchmark} and computational performance in section~\ref{ssec:computational-efficicency}. In each table, we \textbf{bold} the best model(s) and underline the second best model(s). When we benchmark both OMAT and OMOL heads or tasks for different benchmarks, we bold and underline the best and second best models within each task.

\section{Dispersion corrections (D3)}
We incorporate D3(BJ) dispersion corrections~\cite{Grimme2010,Grimme2011} when evaluating systems with significant van der Waals interactions with PBE-trained models, using the torchDFTD3 Python package~\cite{takamoto2021pfp} with the PBE parametrization. All D3(BJ) evaluations use the same parameterisation across models to ensure fair comparison. The models trained on OMOL at the $\omega$B97M-VV10 level of theory were run without additional dispersion correction.

\section{Materials Benchmarks}
\label{sec:materials_benchmarks}

We evaluate model performance across comprehensive benchmarks covering inorganic materials properties, including elastic moduli, thermal conductivity, and phonon spectra. These benchmarks assess the models' ability to predict basic physical properties for inorganic materials. These metrics go beyond just energy and force errors by probing the MLIPs’ understanding of the curvature of the PES.
Note that most reference data for these benchmarks are computed at the PBE+$U$ level, matching the MPtraj and OMAT dataset specifications. 

\subsection{Elastic Moduli}

We benchmark bulk ($B$) and shear ($G$) moduli against DFT PBE and PBE$+U$ references to evaluate the models' ability to capture potential energy surface curvature and mechanical response under small strains.

We use \textbf{MatCalc}'s \textbf{ElasticityCalc}~\cite{Liu_MatCalc_2024} to deform the structures with normal (diagonal) strain magnitudes of
$\pm 0.01$ and $\pm 0.005$ for
$\epsilon_{11}$, $\epsilon_{22}$, $\epsilon_{33}$, and off-diagonal strain magnitudes of $\pm 0.06$ and $\pm 0.03$ for $\epsilon_{23}$, $\epsilon_{13}$, $\epsilon_{12}$. The stress tensor $\boldsymbol{\sigma}$ is calculated for each applied strain $\epsilon_{ij}$ and the resulting sets of stress-strain pairs are used in linear regression to obtain the elastic tensor~\cite{de2015charting}.
Then, the bulk and shear moduli are obtained from the elastic and stress tensors via the Voigt-Reuss-Hill (VRH) average, which combines the upper (Voigt) and lower (Reuss) bounds of the moduli for a more accurate approximation~\cite{chung1967voigt, hill1952elastic, golesorkhtabar2013elastic}.

\begin{table}[h!]
\centering
\caption{Bulk and shear moduli benchmark results. Materials Project elasticity dataset containing 12,122 materials (excluding $B$ or $G < -50$ GPa and $B$ or $G > 600$ GPa). Both the initial and deformed structures were relaxed.}
\label{tab:moduli}
\begin{tabular}{@{}lcc@{}}
\toprule
\textbf{Model} &  B MAE (GPa) & G MAE (GPa) \\
\midrule
mace-mh-1-omat & 12.49  & \textbf{7.95}   \\
mace-omat-1 & 11.50 & 8.09 \\
mace-mp-0a & 11.02 & 20.86 \\
mace-omat-0 & 12.47 & 8.95 \\
mattersim-5M & \underline{10.47} & 9.69 \\
orb-v3-consv-inf-omat & \textbf{7.18} & \underline{8.03} \\
uma-m-1p1-omat & 13.60 & 9.65 \\
uma-s-1p1-omat & 14.33 & 8.18 \\
\midrule
mace-mh-1-omol-1\% &  18.00 & 10.77  \\
uma-m-1p1-omol-100\% &   Not converged & Not converged \\
uma-s-1p1-omol-100\% &   Not converged & Not converged \\
\bottomrule
\end{tabular}
\end{table}

Table~\ref{tab:moduli} reports mean absolute errors for predicted bulk and shear moduli against Materials Project elasticity dataset references~\cite{de2015charting}. The \textbf{orb-v3-consv-inf-omat} model achieves the lowest bulk modulus error, while \textbf{mace-mh-1-omat} achieves the lowest shear modulus error. Overall, the \textbf{mace-mh-1-omat} model demonstrates competitive performance, with accuracy comparable to the backbone OMAT model, confirming that multi-head replay fine-tuning effectively retains bulk property accuracy.

\subsection{Thermal Conductivity}

Thermal conductivity represents a critical property for electronics, thermoelectrics, and energy storage applications. We employ a comprehensive thermal conductivity benchmark~\cite{póta2025thermalconductivitypredictionsfoundation, Riebesell2025} evaluating both microscopic anharmonic phonon properties and resulting lattice thermal conductivity, capturing particle-like (Boltzmann transport equation) and wave-like (Wigner) heat-transport mechanisms across 103 chemically and structurally diverse solids at near first-principles accuracy.

The benchmark encompasses 103 binary crystals—rock salt, zinc-blende, and wurtzite phases spanning 34 chemical elements—with accompanying first-principles harmonic and anharmonic force-constant data enabling reference Wigner-transport conductivities and mode-resolved metrics for rigorous quantitative comparison.

\begin{table}[h!]
\centering
\caption{Thermal conductivity benchmark performance}
\label{tab:thermal_cond}
\begin{tabular}{@{}lc@{}}
\toprule
\textbf{Model} & $\kappa_{\mathrm{RMSE}}$ (W/mK) \\
\midrule
mace-mh-1-omat & 0.24 \\
mace-omat-1 & \underline{0.20} \\
mace-mp-0a & 0.62 \\
mace-omat-0 & 0.24 \\
mattersim-5M & 0.57 \\
orb-v3-consv-inf-omat & 0.21 \\
uma-m-1p1-omat & \textbf{0.17}  \\
uma-s-1p1-omat & \underline{0.20} \\
\bottomrule
\end{tabular}
\end{table}

Table~\ref{tab:thermal_cond} presents root-mean-square errors for predicted lattice thermal conductivity on the 103-compound Wigner-transport benchmark. The \textbf{uma-m-1p1-omat} model achieves the lowest $\kappa_{\mathrm{RMSE}}$, closely followed by \textbf{mace-omat-1} and \textbf{uma-s-1p1-omat}. Our \textbf{mace-mh-1-omat} model maintains strong performance compared to the baseline, confirming robust bulk property prediction.

\subsection{Phonons}

We evaluate phonon frequencies and derived thermodynamic properties using the Materials Data Repository (MDR) phonon benchmark~\cite{Loew2025}, testing approximately 10,000 materials against reference PBE+$U$ calculations. We assess maximum, average, and minimum phonon frequencies, plus mean absolute error (MAE) across the Brillouin zone. Additional thermodynamic properties include entropy ($S$), Helmholtz free energy ($F$), and heat capacity at constant volume ($C_V$) at 300 K, all derived from phonon frequencies.
Structures are relaxed with fixed symmetries matching DFT references, and phonon frequencies are computed using finite difference methods with identical displacement parameters.
For the ``no-sym'' pipeline, we deliberately \emph{do not} apply symmetry fixing because we observed numerical instabilities in finite-displacement workflows when enforcing symmetry constraints for these models; all ``no-sym'' results are therefore obtained without symmetry restoration. The full set of results showing all models with and without symmetry constraints applied is shown in Table~\ref{tab:mdr_benchmark_full}.
\begin{figure}[h!]
    \centering
    \includegraphics[width=0.9\linewidth]{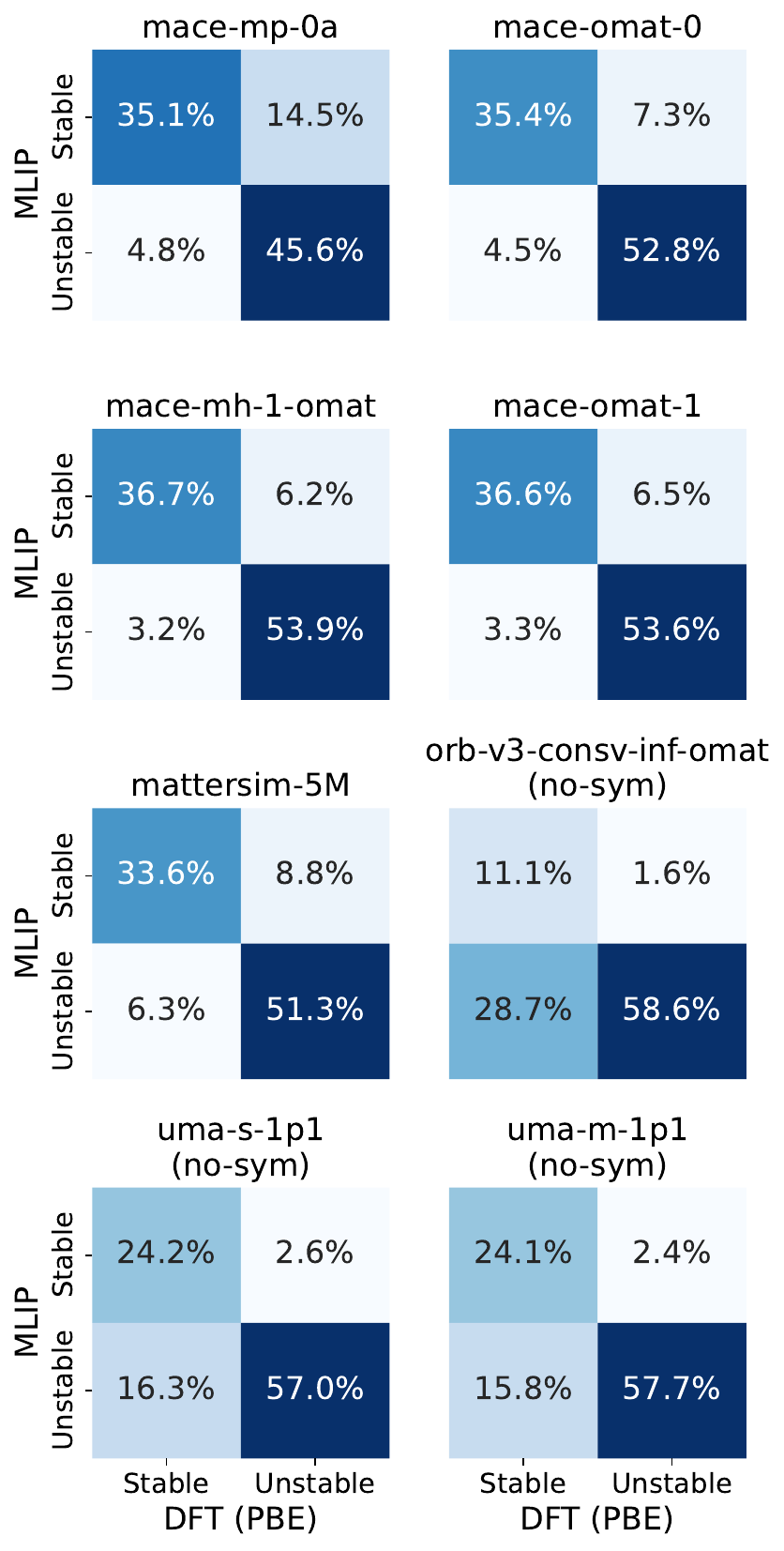}
    \caption{\textbf{Phonon dynamical stability classification confusion matrices.} Materials are classified as unstable if $|\omega_{\mathrm{imag}}| > 0.05\,\text{THz}$ 
    ($\approx 2.4$ K).}
    \label{fig:phonon_stability_cm}
\end{figure}
\begin{table}[h!]
\centering
\caption{MDR Phonon benchmark on the phonon frequencies and thermodynamic properties (300 K) of roughly ten thousand materials. Each column corresponds to the MAE of the named quantity. BZ refers to the MAE across the whole Brillouin Zone.}
\label{tab:mdr_benchmark}
\resizebox{\columnwidth}{!}{%
\begin{tabular}{@{}lccccccc@{}}
\toprule
\textbf{Model} & $\omega_{\text{max}}$ & $\omega_{\text{avg}}$ & $\omega_{\text{min}}$ & BZ & $S$ & $F$ & $C_V$ \\
& (K) & (K) & (K) & (K) & (J/mol$\cdot$K) & (kJ/mol) & (J/mol$\cdot$K) \\
\midrule
mace-mh-1-omat & 12  & \textbf{3} & \textbf{11} & \textbf{5} & \underline{8} & \textbf{2} & \underline{3}  \\
mace-omat-1 & 13  & \textbf{3} & \underline{12} & 8 & \underline{8} & \textbf{2} & \underline{3} \\
mace-mp-0a     & 65  & 32 & 19 & 33 & 60 & 23 & 14 \\
mace-omat-0      & 16 & 4 & 13 & \underline{7} & 10 & 3 & \underline{3} \\
mattersim-5M   & 19  & 5 & 16 & 10 & 14 & 4 & 4  \\
orb-v3-consv-inf-omat (no-sym)   & 12  & 5 & 29 & 15 & 13 & 3 & 4  \\
uma-m-1p1-omat (no-sym)    & \textbf{9}  & \textbf{3} & 18 & 8 & \underline{8} & \textbf{2} & \textbf{2}  \\
uma-s-1p1-omat (no-sym)    & \underline{11}  & 4 & 21 & 9 & \textbf{7} & \textbf{2} & \underline{3}  \\
\midrule
mace-mh-1-omol-1\% & \textbf{49}  & \textbf{12} & \textbf{18} & \textbf{16} & \textbf{16} & \textbf{7} & \textbf{7} \\
uma-s-1p1-omol-100\% (no-sym) & 155  & 50 & 64 & 64 & 82 & 32 & 19 \\
\bottomrule
\end{tabular}%
}
\end{table}

Table~\ref{tab:mdr_benchmark} shows mean absolute errors for phonon frequencies and thermodynamic properties. The \textbf{mace-mh-1-omat} model achieves the lowest errors for most properties, only ranked second best behind \textbf{uma-m-1p1-omat} for $C_V$, third behind \textbf{uma-s-1p1-omat} and \textbf{uma-m-1p1-omat} for $\omega_{\text{max}}$ and second behind \textbf{uma-s-1p1-omat} for $S$, demonstrating excellent phonon frequency and thermodynamic property prediction. Notably, symmetry fixing significantly impacted \textbf{uma} and \textbf{orb-v3} models’ performance while leaving other models unaffected. Our \textbf{mace-mh-1-omat} model maintains accuracy or slightly outperforms the backbone model \textbf{mace-omat-1}, confirming the effectiveness of retaining high bulk materials accuracy through multi-head replay fine-tuning. In Figure~\ref{fig:phonon_stability_cm}, we report dynamical stability classification performance, showing that our \textbf{mace-mh-1-omat} achieves the best overall classification performance. We also compare the OMOL head of \textbf{mace-mh-1} and of \textbf{uma-s-1p1} to further assess cross-learning in multidomain models. We observe a clearly superior degree of cross-learning for \textbf{mace-mh-1} than \textbf{uma-s-1p1}, with the mace model achieving a phonon performance superior in each domain, even outperforming the specialised MPTraj model \textbf{mace-mp-0a}. The \textbf{uma-m-1p1-omol} was unable to geometry optimise a large portion of the structures, so we did not include it in the table.

\section{Molecular Crystal Benchmarks}
\label{sec:molecular_crystals}

Molecular crystal formation energies test the models' accuracy in describing intermolecular interactions and the cohesive assembly of molecular solids.

\subsection{X23 Molecular Crystals}

The X23 benchmark~\cite{reilly2013understanding} comprises 23 experimentally characterised organic molecular crystals selected to span diverse noncovalent binding motifs, including hydrogen bonding, dispersion-dominated packing, and mixed electrostatic–van der Waals interactions. Reference formation enthalpies are computed via periodic calculations using Diffusion Monte Carlo (DMC)~\cite{DellaPia2024}. Comparing per-molecule cohesive energies against these high-level references evaluates the ability to capture subtle intermolecular forces governing crystal stability.
\begin{table}[h!]
  \centering
  \caption{X23 benchmark formation energy mean absolute errors.}
  \label{tab:x23}
  \begin{tabular}{@{}lc@{}}
    \toprule
    \textbf{Model}                   & MAE (kJ/mol) \\
    \midrule
    mace-mh-1-omat-D3 & \textbf{15.82} \\
    mace-omat-1-D3 & 19.60 \\
    mace-mp-0a-D3 & \underline{17.42} \\
    mace-omat-0-D3 & 53.23   \\
    mattersim-5M-D3 & 20.11 \\
    orb-v3-consv-inf-omat & 28.76 \\
    uma-m-1p1-omat-D3 & 78.89   \\
    uma-s-1p1-omat-D3 & 27.99   \\
    \midrule
    mace-mh-1-omol-1\% & \underline{7.41} \\
    uma-m-1p1-omol-100\% & 9.33  \\
    uma-s-1p1-omol-100\% &  \textbf{6.81}  \\
    \bottomrule
  \end{tabular}
\end{table}

Table~\ref{tab:x23} shows that \textbf{mace-mh-1-omat-D3 } achieves the lowest MAE among the tested models for the X23 benchmark, indicating improved molecular crystal formation energy prediction. Notably, it outperforms \textbf{uma-s-1p1-omat-D3}, despite the latter being trained on larger datasets including 20M molecular crystals. The substantial improvement over \textbf{mace-omat-1-D3} demonstrates clear knowledge transfer of molecular chemistry from OMOL and SPICE heads to the PBE head, transferring molecular chemical knowledge. We observe that the OMOL-trained models outperform the PBE models, which is likely due to a much better description of intermolecular interaction by the OMOL range-separated hybrid DFT $\omega$B97M-VV10.

\subsection{DMC Water Ice Polymorphs}

We test model performance on formation energies of common ice phases (Ih, II, III, etc.), benchmarking hydrogen-bonded networks under different pressures~\cite{della2022dmc}.

\begin{table}[h!]
\centering
\caption{Ice phase mean absolute errors}
\label{tab:ice}
\begin{tabular}{@{}lc@{}}
\toprule
\textbf{Model} & MAE (meV) \\
\midrule
mace-mh-1-omat-D3 & \textbf{11.23} \\
mace-omat-1-D3 & 144.57 \\
mace-mp-0a-D3 & \underline{59.79} \\
mace-omat-0-D3 & 447.08   \\
mattersim-5M-D3 & 96.95 \\
orb-v3-consv-inf-omat & 138.44 \\
uma-m1-p1-D3 & 545.72 \\
uma-s1-p1-D3 & 310.82 \\
\midrule
mace-mh-1-omol-1\% & \textbf{79.60} \\
uma-m-1p1-omol-100\% & 120.62 \\
uma-s-1p1-omol-100\% & \underline{109.20} \\
\bottomrule
\end{tabular}
\end{table}

Table~\ref{tab:ice} demonstrates that \textbf{mace-mh-1-omat-D3} achieves the lowest MAE for ice polymorph relative energies among the tested models,, indicating improved modelling of hydrogen bonding interactions in this benchmark. The significant improvement over \textbf{mace-omat-1-D3} again demonstrates effective fine-tuning enhancement for molecular crystal systems and better description of subtle intermolecular interactions. Surprisingly, the \textbf{mace-mh-1-omat-D3} even outperforms the omol-trained models.

\section{Surface Benchmarks}
\label{sec:surface_benchmarks}

Surface interaction benchmarks probe adsorption energetics and site-specific binding, critical for catalysis and gas–surface processes.

\subsection{General Surface Adsorption: S24 Dataset}

Accurately describing molecule–surface interactions at the first-principles level is essential for designing advanced catalysts, gas-separation membranes, and sensing materials. We employ the S24 benchmark set~\cite{batatia2024foundationmodelatomisticmaterials}, covering 24 prototypical adsorption systems across diverse surface types:
\begin{itemize}
    \item \textbf{Covalent surfaces}: graphene, silicene, and other 2D covalent networks
    \item \textbf{Ionic surfaces}: alkali halide (e.g., NaCl) and metal oxide (e.g., MgO) slabs
    \item \textbf{Metallic facets}: close-packed (111) and stepped surfaces of Pt, Au, and Cu
    \item \textbf{Porous materials}: representative metal–organic frameworks (MOFs) and zeolites
\end{itemize}

Reference adsorption energies are computed at PBE+D3(BJ) level using standardised VASP/MPRelaxSet protocols. We report mean absolute deviations for predicted versus reference energies (in eV) both overall and per surface category.

\begin{table}[h!]
\centering
\caption{S24 benchmark mean absolute errors for the adsorption energies of small molecules on surfaces and porous materials.}
\begin{tabular}{lc}
\toprule
\textbf{Model} & MAE (eV) \\
\midrule
mace-mh-1-omat-D3 & \textbf{0.095} \\
mace-omat-1-D3 & \underline{0.140} \\
mace-mp-0a-D3 & 0.152 \\
mace-omat-0-D3 & 0.550 \\
mattersim-5M-D3 & 0.141 \\
orb-v3-consv-inf-omat-D3 & 0.174 \\
uma-m-1p1-omat-D3 & 0.523 \\
uma-s-1p1-omat-D3 & 0.329 \\
\midrule
mace-mh-1-omol-1\% & \textbf{0.288}  \\
uma-m-1p1-omol-100\% & 10.774 \\
uma-s-1p1-omol-100\% &  \underline{4.636}  \\
\bottomrule
\end{tabular}
\label{tab:mae-eval}
\end{table}

Table~\ref{tab:mae-eval} shows that \textbf{mace-mh-1-omat-D3} achieves the lowest MAE for surface adsorption energies, demonstrating a good description of interactions of molecules with surfaces and porous materials. In particular, we observe a 68\% improvement compared to the backbone model \textbf{mace-omat-1-D3} on the adsportion energies. When testing the OMOL tasks of the models, we observe much better performance for the \textbf{mace-mh-1-omol-1\%} compared to the \textbf{uma} models confirming superior cross-learning on this benchmark. 

\subsection{OC20 Metal Surface Benchmark}

We benchmark small-molecule adsorption on metal surfaces using structures generated during the Open Catalyst Challenge 2023~\cite{chanussot2021open, tran2023open} to evaluate catalytic reaction intermediate modelling. We use the set of configurations from the Open Catalyst Challenge 2023 recomputed at the MPtraj DFT level introduced in Ref.~\cite{batatia2024foundationmodelatomisticmaterials}.

\begin{table}[h!]
\centering
\caption{Open Catalyst Challenge 2023 generated set of adsorption energies of small molecules on metallic surfaces.}
\label{tab:oc20}
\begin{tabular}{@{}lcc@{}}
\toprule
\textbf{Model} & MAE (eV) & Pearson’s r \\
\midrule
mace-mh-1-omat-D3 & \underline{0.138} & \underline{0.98} \\
mace-omat-1-D3 & 0.171 & 0.97 \\
mace-mp-0a-D3 & 0.412 & 0.86 \\
mace-omat-0-D3 & 0.243 & 0.94 \\
mattersim-5M-D3 & 0.285 & 0.92 \\
orb-v3-consv-inf-omat-D3 & 0.159 & 0.974 \\
uma-m-1p1-omat-D3 & \textbf{0.097} & \textbf{0.99} \\
uma-s-1p1-omat-D3 & 0.172 & 0.97 \\
\midrule
mace-mh-1-omol-1\% & \textbf{0.214} & \textbf{0.96} \\
uma-m-1p1-omol-100\% & 3.394 & 0.28 \\
uma-s-1p1-omol-100\% & \underline{1.120} & \underline{0.50} \\
\bottomrule
\end{tabular}
\end{table}

We observe that \textbf{uma-m-1p1-omat-D3} achieves the best performance with an MAE of 0.097 eV followed by \textbf{mace-mh-1-omat-D3} with an MAE of 0.138 eV. Note that \textbf{uma-m-1p1-omat-D3} and \textbf{uma-s-1p1-omat-D3} are trained on the full OC20 dataset representing 100 times more surface configurations than our \textbf{mace-mh-1-omat-D3} model. When benchmarking the OMOL tasks, we confirm the better cross-learning of the \textbf{mace-mh} model compared to the \textbf{uma} models.

\section{Molecular Benchmarks}
\label{sec:molecular_benchmarks}

We evaluate model performance across comprehensive molecular property benchmarks, including conformer energies, reaction energies, and noncovalent interactions, assessing capability to predict key chemical properties governing molecular behaviour.

\subsection{Wiggle 150}
Wiggle150 is a benchmark comprising 150 highly strained conformations of adenosine, benzylpenicillin, and efavirenz. 
Table~\ref{tab:wiggle150} reports mean absolute errors for strained conformer relative energies against Wiggle150 benchmark references~\cite{brew2025wiggle150} computed at the DLPNO-CCSD(T)/CBS level of theory.

\begin{table}[h!]
\centering
\caption{Wiggle150 strained conformer relative energy benchmark with D3(BJ) dispersion correction}
\label{tab:wiggle150}
\begin{tabular}{@{}lc@{}}
\toprule
\textbf{Model} & MAE (kcal\,mol$^{-1}$) \\
\midrule
$\omega$B97M-D3 & 1.18 \\
PBE-D3 & 4.91 \\
\midrule
mace-mh-1-omat-D3 & \textbf{4.80} \\
mace-omat-1-D3 & 10.39 \\
mace-mp-0a-D3 & 25.90 \\
mace-omat-0-D3 & 9.62 \\
mattersim-5M-D3 & 12.12 \\
orb-v3-consv-inf-omat-D3 & 7.65 \\
uma-m-1p1-omat-D3 & \underline{5.04} \\
uma-s-1p1-omat-D3 & 6.60 \\
\midrule
mace-mh-1-omol-1\% & 1.30 \\
mace-omol-100\% & \textbf{0.83} \\
uma-m-1p1-omol-100\% &  0.93  \\
uma-s-1p1-omol-100\% &  \underline{0.91}  \\
\bottomrule
\end{tabular}
\end{table}

Our \textbf{mace-mh-1-omat-D3} model achieves 4.80 kcal\,mol$^{-1}$ MAE, comparable to the underlying PBE-D3 functional performance. The multi-head fine-tuning improves by a factor of 2 over the pre-trained model's baseline (\textbf{mace-omat-1-D3}).
 
\subsection{GMTKN55 Main Group Chemistry}

We evaluate gas-phase chemical accuracy using the GMTKN55 benchmark suite~\cite{goerigk2017look}, which represents the most comprehensive test of electronic structure methods for main group thermochemistry, kinetics and noncovalent interactions. The complete suite comprises 55 individual test sets totalling 1,505 relative energies, systematically categorized into five chemical domains:

\begin{enumerate}
\item \textbf{Basic Properties (8 sets)}: Atomic energies, ionization potentials, electron affinities
\item \textbf{Reaction Energies (15 sets)}: Thermochemistry including G2/97, G3/99 test sets
\item \textbf{Barrier Heights (12 sets)}: Transition state energetics for fundamental organic reactions
\item \textbf{Intramolecular Noncovalent (7 sets)}: Conformational energetics, hydrogen bonding
\item \textbf{Intermolecular Noncovalent (13 sets)}: Dimers, clusters, host-guest complexes
\end{enumerate}

All reference energies employ high-level coupled cluster methods, primarily CCSD(T) extrapolated to the complete basis set limit with core correlation corrections where appropriate. The weighted total mean absolute deviation (WTMAD) provides a single metric combining performance across all chemical domains, with typical values of 2-3 kcal\,mol$^{-1}$ representing chemical accuracy for electronic structure methods. We keep only the neutral, singlet subset as most of the models do not handle charged systems yet.

\begin{table}[h!]
  \caption{\label{tab:gmtkn_comprehensive}GMTKN55 subset mean absolute errors across benchmark categories (neutral singlet subset only). All values represent weighted mean absolute errors in kcal\,mol$^{-1}$ (lower is better) with weights taken from WMAD-2.}
  \resizebox{\columnwidth}{!}{%
    \begin{tabular}{lcccccc}
      \hline
      \textbf{Model} & Basic & Large & Barrier & Intra- & Inter- & All \\
            & Small & Systems & Heights & Noncov & Noncov & \\
      \hline
      $\omega$B97M-D3BJ                 & 2.86	& 5.77 & 2.34	& 4.54 & 3.63 & 4.04 \\
      PBE-D3(BJ)                 & 11.24 & 10.61 & 13.16 &  9.95 &  9.89 & 10.58 \\
      \midrule
      mace-mh-1-omat-D3  & \textbf{12.88}  & \textbf{13.74} & \textbf{9.58} & \textbf{11.45} & \textbf{9.54} & \textbf{11.23} \\
      mace-omat-1-D3  & 25.05  & 33.65 & 18.95 & \underline{26.47} & \underline{20.59} & 24.90 \\
      mace-mp-0a-D3        & 42.62 & 59.75 & 31.41 & 59.60 & 23.27 & 45.04 \\
      mace-omat-0-D3        & 40.61 & 44.51 & 30.49 & 57.73 &111.69 & 64.15 \\
      mattersim-5M-D3      & 29.27 & 41.36 & 20.34 & 37.40 & 23.50 & 31.46 \\
      orb-v3-consv-inf-omat-D3 & 26.30 & \underline{14.72} & \underline{13.88} & 28.09 & 23.08 & \underline{22.30} \\
      uma-m-1-p1-omat-D3       & \underline{25.02} & 20.77 & 28.34 & 23.33 & 138.21 & 52.89 \\
      uma-s-1-p1-omat-D3       & 29.35 & 28.31 & 15.24 & 33.64 & 36.75 & 30.83 \\
      \hline
      AIMNet2                  & 11.91 & 18.67 & 10.80 & 12.19 & 20.76 & 15.15 \\
      MACE-OFF23(L)            &  \underline{8.74} & 10.40 & 22.52 &  6.79 & 37.84 & 16.46 \\
      mace-mh-1-omol-1\% & 10.62 & 7.64  & 7.10 & 7.47 & 10.11 & 8.44   \\
      mace-omol-100\% & 9.41 & 4.33&  3.06 & \textbf{4.96} & 3.59 & 4.65  \\
      uma-m-1p1-omol-100\% & 12.18 & \underline{3.93} & \underline{2.91} & \textbf{4.96} & \textbf{2.34} & \underline{4.49}     \\
      uma-s-1p1-omol-100\% & \textbf{8.23} & \textbf{3.50} & \textbf{2.76} & 5.15 & \underline{2.86} & \textbf{4.22}   \\
      \hline
    \end{tabular}%
  }
\end{table}
Table~\ref{tab:gmtkn_comprehensive} shows that \textbf{mace-mh-1-omat-D3} significantly outperforms other PBE-trained foundational models, including \textbf{uma-m-1-p1-omat-D3} achieving amean error on the GMTKN55 of 11.23 kcal\,mol$^{-1}$, comparable to the accuracy of PBE-D3(BJ). The substantial improvement over \textbf{mace-omat-1-D3} confirms the effective knowledge transfer from the molecular heads (SPICE and OMOL) to the material head. Notably, our model achieves performance comparable to specialised molecular models (AIMNet2, MACE-OFF23(L)) that were trained on hybrid meta-GGA functional $\omega$B97M-D3BJ, which is much more accurate for molecular system.

\subsection{Protein fragments: PLF547}

Accurately capturing intramolecular hydrogen bonding and side–chain/backbone contacts in polypeptide fragments is a stringent test for any MLIP intended to model biomolecular chemistry. We therefore evaluate on the PLF547 benchmark suite~\cite{kriz2020benchmarking} of 547 shorter peptide-like fragments (PLF547) with interaction energies referenced to high-level DLPNO-CCSD(T)/CBS calculations. Following prior work, we compute single-point energies on the published geometries and report the mean absolute error between the predicted and reference interaction energies, which provides a good proxy to the quality of intermolecular interactions. We used only the subset of molecules that are neutral singlet.

\begin{table}[h!]
\centering
\caption{PLF547 neutral subset interaction energy MAE (kcal\,mol$^{-1}$) for the different tested models.}
\label{tab:PLF}
\begin{tabular}{@{}lcc@{}}
\toprule
\textbf{Model} & PLF547 MAE (kcal\,mol$^{-1}$)\\
\midrule
mace-mh-1-omat-D3 & \textbf{0.626} \\
mace-omat-1-D3 & \underline{0.839}\\
mace-mp-0a-D3 & 1.040\\
mace-omat-0-D3 & 4.926\\
mattersim-5M-D3 & 1.017\\
orb-v3-consv-inf-omat-D3 & 1.829 \\
uma-m-1p1-omat-D3 & 12.057 \\
uma-s-1p1-omat-D3 & 2.935\\
\midrule
mace-omol-100\% & \textbf{0.334} \\
mace-mh-1-omol-1\% & \underline{0.394} \\
uma-m-1p1-omol-100\% & 0.839 \\
uma-s-1p1-omol-100\% & 0.655  \\
\bottomrule
\end{tabular}
\end{table}

Table~\ref{tab:PLF} summarizes the results. The \textbf{mace-mh-1-omat-D3} model achieves the lowest MAE, with 0.626 kcal\,mol$^{-1}$, substantially improving over the OMAT-only backbone. Overall, these results demonstrate that cross-domain replay  enhances the model's ability to describe biomolecular fragment energetics. 

\subsection{S30L Molecular complexes}

Large host–guest and $\pi - \pi$ stacked complexes probe the long-range dispersion and subtle many-body polarization effects that drive supramolecular binding. We assess these regimes with the S30L benchmark~\cite{sure2015comprehensive}, a set of 30 noncovalent complexes ranging from crown-ether inclusion compounds to charged receptor–ligand pairs. Reference binding energies are empirical binding energies obtained by back-correcting the experimental association free energies, making S30L a challenging benchmark for dispersion-dominated interactions.

\begin{table}[h!]
\centering
\caption{S30L~\cite{sure2015comprehensive} benchmarking of host-guest binding energies in large molecular complexes. Binding energies mean absolute errors in kcal\,mol$^{-1}$ compared to experimental references.}
\label{tab:S30L}
\begin{tabular}{@{}lc@{}}
\toprule
\textbf{Model} & MAE (kcal\,mol$^{-1}$)\\
\midrule
mace-mh-1-omat-D3 & \textbf{10.13} \\
mace-omat-1-D3 & 15.35 \\
mace-mp-0a-D3 & 14.22 \\
mace-omat-0-D3 & 25.87   \\
mattersim-5M-D3 & \underline{11.92} \\
orb-v3-consv-inf-omat-D3 & 13.64 \\
uma-m-1p1-omat-D3 & 13.09 \\
uma-s-1p1-omat-D3 & 15.14 \\
\midrule
mace-mh-1-omol-1\% &  \textbf{6.66}  \\
uma-m-1p1-omol-100\% &  \underline{7.66} \\
uma-s-1p1-omol-100\% &  14.59  \\
\bottomrule
\end{tabular}
\end{table}

We evaluate single-point interaction energies on the published geometries and report mean absolute errors (MAE) in kcal\,mol$^{-1}$ (Table~\ref{tab:S30L}). The \textbf{mace-mh-1-omat-D3} model attains the best performance, with an MAE of 10.13 kcal\,mol$^{-1}$, substantially outperforming pre-trained model \textbf{mace-omat-1-D3} at 15.35 kcal\,mol$^{-1}$. In contrast, UMA models—despite training on vastly more molecular data—achieve larger errors, reinforcing that architectural choices on how to merge different datasets are critical for enhancing cross-learning behaviour. Note also that the total charge information was not given to the UMA OMAT task models as it was causing very large errors, which highlights that naive inclusion of total charge via global embedding is not transferable between the models' tasks.

\section{Physicality Benchmarks}
\label{sec:physicality_benchmark}

While extensive accuracy benchmarks are essential, ensuring physically realistic predictions is equally important, given the models' broad applicability across chemical domains that makes exhaustive validation challenging. We evaluate model physicality through several benchmarks assessing size extensivity, additivity, and smoothness of pair interactions.

\subsection{Size Extensivity and Locality}

The size extensivity and locality of the potential energy surface represent fundamental quantum mechanical properties, ensuring that the energy scales correctly with particle number and that sufficiently separated system energies equal their component sums. Large violations of these properties can lead to unphysical simulations.
\begin{table}[h!]
\centering
\caption{Slab test for size extensivity evaluation. $\Delta$ refers to the difference between the isolated slab energies and the combined system energy ($\Delta = E_{1,2} - (E_1 +E_2)$). Results to 1 d.p.}
\label{tab:extensivity_slabs}
\resizebox{\columnwidth}{!}{%
\begin{tabular}{@{}lcccc@{}}
\toprule
\textbf{Model} & $E_1$ & $E_2$ & $E_{12}$ & $\Delta$ \\
      & (eV)  & (eV)  & (eV)     & (meV)    \\
\midrule
mace-mh-1-omat & -468.2 & -615.0 & -1083.3 & \textbf{0.0} \\
mace-omat-1 & -467.3 & -617.0 & -1084.3 & \textbf{0.0} \\
mace-mp-0a     & -468.1 & -649.5 & -1117.6 & \textbf{0.0} \\
mace-omat-0    & -468.8 & -616.0 & -1084.7 & \textbf{0.0} \\
mattersim-5M   & -467.8 & -646.7 & -1114.5 & 0.2 \\
orb-v3-consv-inf-omat & -466.3 & -614.1 & -1081.1 & -709.7 \\
uma-m-1p1-omat    & -467.2 & -616.4 & -1081.2 & 1436.9 \\
uma-s-1p1-omat    & -467.4 & -616.6 & -1084.5 & -453.8 \\
\midrule
mace-mh-1-omol-1\% & -844543.4 & -5253593.4 & -6098136.8 & \textbf{0.0} \\
uma-m-1p1-omol-100\% & -843663.9 & -5253001.4 & -6097051.2 & -385824.5  \\
uma-s-1p1-omol-100\% & -843500.5 & -5253244.0 & -6096655.8 & 88670.6  \\
\bottomrule
\end{tabular}}
\end{table}

For the slab test, we compare the sum of the energies of isolated FCC(111) aluminium ($E_1$) and nickel ($E_2$) slabs, each 8 layers thick in a 4x4 in-plane supercell, against the energy of the combined configuration ($E_{12}$), with a 100 \AA\ gap. The expected non-interacting behaviour is confirmed if $E_{12} = E_1 + E_2$.

Table~\ref{tab:extensivity_slabs} shows all models except UMA models, and ORB-V3-Consv-Inf-omat maintain proper extensivity, while UMA models and ORB models exhibit large energy deviations, indicating the presence of unphysical interactions. For the UMA models, this non-local interaction is likely arising from global chemical element embeddings creating non-local interactions. For the ORB model, it is due to its non-local readout in which it passes a summed or averaged energy over the entire structure into a non-linear function, creating potential unphysical non-local interactions.

\begin{table}[h!]
\centering
\caption{Addition of a hydrogen atom to an aluminium slab at 50 \AA\ distance to test for size additivity.}
\label{tab:extensivity_H_atom}
\begin{tabular}{@{}lccc@{}}
\toprule
\textbf{Model} & max $|\Delta F|$ & mean $|\Delta F|$ & std $|\Delta F|$ \\
      & (meV/\AA)         & (meV/\AA)         & (meV/\AA)\\
\midrule
mace-mh-1-omat & \textbf{0.0000} & \textbf{0.0000} & \textbf{0.0000} \\
mace-omat-1 & \textbf{0.0000} & \textbf{0.0000} & \textbf{0.0000} \\
mace-mp-0a     & \textbf{0.0000} & \textbf{0.0000} & \textbf{0.0000} \\
mace-omat-0    & \textbf{0.0000} & \textbf{0.0000} & \textbf{0.0000} \\
mattersim-5M   & \textbf{0.0000} & \textbf{0.0000} & \textbf{0.0000} \\
orb-v3-consv-inf-omat & 61.65 & 19.21 & \textbf{0.0000} \\
uma-m-1p1-omat    & 1520.0  & 11.73 & 0.0006 \\
uma-s-1p1-omat    & 969.20  & 16.48 & 0.0005 \\
\midrule
mace-mh-1-omol-1\% & \textbf{0.0000} & \textbf{0.0000} & \textbf{0.0000} \\
uma-m-1p1-omol-100\% &  38.17 & 0.227 & 0.0003  \\
uma-s-1p1-omol-100\% &  719.2 & 1.018 & 0.0002  \\
\bottomrule
\end{tabular}
\end{table}

We further test size additivity by placing a hydrogen atom 50 Å from an aluminium slab and computing forces at various distances. Table~\ref{tab:extensivity_H_atom} shows all models except \textbf{uma-s-1p1-omat}, \textbf{uma-m-1p1-omat} and \textbf{orb-v3-consv-inf-omat} produce zero forces as expected for non-interacting systems. The \textbf{uma} models and orb models exhibit large spurious forces, with for example \textbf{uma-s-1p1-omat} showing a spurious force of max 969.2 meV/Å, confirming large unphysical interactions.

\subsection{Homonuclear and Heteronuclear Diatomics}

Diatomic potential curve analysis provides fundamental tests of model smoothness and physicality for low-order body interactions.
\begin{table}[h!]
\centering
\caption{Homonuclear and heteronuclear diatomic physicality assessment across key smoothness and correlation metrics. Force flips count the number of sign changes in the force, with the ideal being a single flip from repulsive to attractive. Energy minima denote the number of distinct local minima in the potential energy curve, with the physical expectation of only one bound state. Energy inflections capture the number of inflections in the potential energy curve, where the ideal curve would have one inflection point. The Spearman's rank correlation coefficients for the repulsive and attractive regimes have ideal values of $\rho_{E_{rep}} = -1$ and $\rho_{E_{at}} = 1$ respectively for perfectly monotonic relationships.}
\label{tab:diatomic_physicality}
\resizebox{\columnwidth}{!}{%
\begin{tabular}{@{}lccccc@{}}
\toprule
\textbf{Model} & \makecell{Mean No. \\ Force Flips} & \makecell{Mean No. \\ Energy Minima} & \makecell{Mean No. \\ Energy Inflections} & \makecell{Mean \\ $\rho_{E_{\text{rep}}}$} & \makecell{Mean \\ $\rho_{E_{\text{at}}}$} \\

\midrule
mace-mh-1-omat      & 2.09          & \underline{1.42}          & 2.72          & \underline{-0.99} & \underline{0.88} \\
mace-omat-1      & \textbf{1.52}          & \textbf{1.18}          & \textbf{2.15}          & \textbf{-1.00} & 0.82 \\
mace-mp-0a        & 3.75          & 2.15          & 3.32          & -0.96          & 0.15 \\
mace-omat-0         & \underline{1.97} & 1.45 & \underline{2.24} & \textbf{-1.00} & 0.78 \\
mattersim-5M      & 2.17          & 1.61          & 2.45          & \textbf{-1.00} & \textbf{0.89} \\
orb-v3-consv-inf-omat  & 2.91          & 1.62          & 2.56          & -0.98 & 0.63 \\
uma-m-1p1-omat    & 8.83          & 4.16          & 12.85         & -0.57          & 0.56 \\
uma-s-1p1-omat    & 10.73         & 4.82          & 15.55         & -0.70          & 0.42 \\
\midrule
mace-mh-1-omol-1\% & \textbf{2.03} & \textbf{3.39} & \textbf{3.70} & \textbf{-0.99} & \textbf{0.87}\\
uma-m-1p1-omol-100\%  & 12.09 & 5.56 & 16.11 & -0.83 & \underline{0.77}  \\
uma-s-1p1-omol-100\% & \underline{11.40} & \underline{5.27} & \underline{16.06} & \underline{-0.93} & 0.76 \\
\bottomrule
\end{tabular}%
}
\end{table}
As diatomic molecules have shown convergence issues for plane-wave DFT reference data, one can instead define a set of general physical requirements for the ideal dimer curve, where the choice of these metrics was inspired by MLIP Arena~\cite{chiang2025mlip}. Such a curve should exhibit a single energy minimum, a single energy inflection, and one change in force sign (force flip). To quantify smoothness and monotonicity, Spearman's rank correlation coefficients are used. The ideal dimer potential energy curve can be decomposed into two monotonic functions of energy and atomic separation either side of the energy minimum, with a repulsive regime at shorter separations and an attractive regime at longer separations.
Table~\ref{tab:diatomic_physicality} evaluates diatomic curve quality using mean values of each metric over all compatible homonuclear and heteronuclear diatomics.
The \textbf{mace-omat-1} and \textbf{mace-mh-1-omat} models demonstrate good performance with minimal force flips and energy minima. However, \textbf{uma-s-1p1-omat(omol)} and \textbf{uma-m-1p1-omat(omol)} exhibit poor diatomic behaviour with numerous force sign flips, energy minima, and inflection points, indicating problematic two-body interaction smoothness. This is confirmed upon visual inspection of the \textbf{uma} models’ diatomic curves. We also attach in the supplementary material compressed files containing all the homonuclear diatomic curves for the models tested in the paper.

\section{Computational efficiency}
\label{ssec:computational-efficicency}

The computational efficiency is assessed by comparing the time required to calculate the energy and forces of 1,000 atoms in Carbon FCC structures with a lattice constant $a=3.8$ \AA~on a single NVIDIA H100 80GB
GPU using FP32 (TF32-high precision). Numbers for UMA models and ORB are taken from~\cite{wood2025umafamilyuniversalmodels}, from which we reproduce the timing protocol. For the new MACE timings, we use both NVIDIA's cuEquivariance kernels~\cite{firoz2025optimizingdatadistributionkernel} and torch.compile with 'reduce-overhead' settings.
\begin{table}[h!]
\centering
\caption{Single GPU (NVIDIA H100 80GB GPU) speed comparison between models to compute energy and forces of a 1000 atoms diamond structure, using torch.compile and for MACE models cuEquivariance~\cite{firoz2025optimizingdatadistributionkernel} kernels, excluding graph construction.}
\label{tab:speed_steps}
\begin{tabular}{@{}lcc@{}}
\toprule
\textbf{Model} & Steps per second \\
\midrule
mace-mh-1 & 43 \\
mace-omat-1 & 43 \\
mace-mp-0a & \textbf{83} \\
mace-omat-0 &  \textbf{83} \\
orb-v3-consv-inf-omat & 30 \\
uma-m-1p1 & 3  \\
uma-s-1p1 & 16 \\
\bottomrule
\end{tabular}
\end{table}

We observe that our models achieve competitive speed compared to the other state-of-the-art models in the tested setup, achieving a speed of 43 steps per second (around 3.8 Megasteps/day) in a best-case scenario (neglecting any molecular dynamics (MD) overheads).  The smaller inference speed of the \textbf{mace-omat-1} and \textbf{mace-mh-1-omat} compared to \textbf{mace-mp-0a} and \textbf{mace-omat-0} is mainly due to a choice of large hyperparameters (L=2 messages and 512 channels for the node features) and not to the architecture changes outlined in Section \ref{sec:non-lin-block}. We note that a fair assessment of the speed of MLIPs is a hard task, that depends on many factors, such as MD drivers, floating-point precisions, compilations and kernels, system density and size, and type of GPU.
\begin{table}[h!]
\centering
\caption{Single GPU (Nvidia H100 80GB GPU) speed comparison between models to run molecular dynamics NVT simulation of a 1000 atoms diamond structure, using MACE models cuEquivariance~\cite{firoz2025optimizingdatadistributionkernel} kernels in LAMMPS MLIAP (no torch.compile).}
\label{tab:speed_md}
\begin{tabular}{@{}lcc@{}}
\toprule
\textbf{Model} & Mega-steps per day \\
\midrule
mace-mh-1 & 1.4 \\
mace-omat-1 & 1.4 \\
mace-mp-0a & 2.2 \\
mace-omat-0 & 2.2 \\
\bottomrule
\end{tabular}
\end{table}

We also benchmark the MACE models during real molecular dynamics in the LAMMPS MLIAP KOKKOS interface of the same carbon structure, using cuEquivariance in float32 but not torch.compile. We observe around 1.4 to 2.2 mega-steps per day of simulation. Compared to the idealised timings of Table~\ref{tab:speed_steps}, this represents a slowdown of around a factor of 2.5, which is mainly due to not using the torch.compile and to a smaller extent, to the various overheads of graph construction and LAMMPS updates. We are working on further engineering optimisations that will enable us to get closer to the ideal models’ speed in real simulations.

\section{Discussion and Outlook}
\label{sec:discussion}

Our results provide strong evidence for effective knowledge transfer across chemical domains through shared representational learning. The substantial improvement observed in molecular systems for the OMAT head when including the SPICE and OMOL datasets as part of the multi-head model demonstrates that molecular knowledge can be transferred to the material head through an improved description of local atomic environments and coordination patterns. 
The multi-head architecture enables knowledge sharing while maintaining consistency across different levels of electronic structure theory in its loss functions. We observe that the more flexible global embedding of the level of theory of UMA does not exhibit a similar level of transfer of knowledge, suggesting that architecture choices and reduced flexibility may enhance efficient transfer..

This work establishes the foundation for next-generation simulation capabilities where single models seamlessly handle complex multiscale phenomena spanning molecular, surface, and materials chemistry. As the field moves toward foundation MLIPs that are out of the box as accurate as a GGA DFTs for most of chemistry with a
single model, the principles demonstrated here provide valuable guidance for achieving both breadth and accuracy in chemical modelling applications.

Several promising extensions emerge from this work: (i) incorporation of additional chemical domains including solid/liquid interfaces~\cite{sahoo2025opencatalyst2025oc25}, molecular crystals~\cite{gharakhanyan2025openmolecularcrystals2025}, or amorphous systems, (ii) development of uncertainty quantification methods for reliable out-of-domain predictions, (iii) integration with experimental data through hybrid learning approaches, and (iv) extension to charged and magnetic systems with the inclusion of electrostatic interactions and spin states (available in some datasets) that will further enhance both accuracy and transferability.

\section*{Supplementary Material}

The supplementary material contains details of the hyperparameters of the models, weighting schemes for the benchmark scores, and additional R2SCAN model results. We also attach in the supplementary material compressed files containing all the homonuclear diatomic curves for the models tested in the paper.

\section*{Conflict of Interest}
GC is a partner in Symmetric Group LLP that licenses force fields commercially and also has equity interest in Ångström AI. SWN has financial interest and equity stake in Mirror Physics, a company working on AI and atomistic modelling.

\section*{Data and Models Availability}
The MACE code is available on \url{https://github.com/ACEsuit/mace} and example input scripts and pre-trained models to reproduce the results are provided on the MACE foundation GitHub: \url{https://github.com/ACEsuit/mace-foundations}. The datasets used for training are all public and referenced in the text. Processed data and analysis scripts to reproduce the benchmarks will be made available in an upcoming publication on ML Potential Usability and Performance Guide
(ML-PEG) \url{https://github.com/ddmms/ml-peg} and live \url{http://ml-peg.stfc.ac.uk}.
The cuEquivariance kernels for MACE are available here: \url{https://github.com/NVIDIA/cuEquivariance}.

\section*{Acknowledgments}
We would like to thank Domantas Kuryla for providing the RGD1 dataset that he recomputed at the B3LYP/6-31G* level of theory. J. H. would like to thank Bal\'azs P\'ota for discussions on phonons.
We acknowledge the Jean Zay cluster access to compute as part of the Grand Challenge: GC010815458 (Grand Challenge Jean Zay H100). We would like to thank the Jean Zay cluster team and administration, as well as GENCI, for the continual help in using the Jean Zay cluster. We would like to thank the Max Planck Computing and Data Facility for providing access to the Raven HPC system, which enabled the computation of many benchmarks. We would like to thank Sovereign AI and Isambard-AI for providing additional compute to run experiments.
We are grateful for
computational support from the UK national high-performance computing service,
ARCHER2, for which access was obtained via the UKCP consortium and funded by
EPSRC grant reference EP/P022065/1 and EP/X035891/1.
I.B. was supported by the Harding Distinguished Postgraduate Scholarship. J. H. was supported by The Lennard-Jones Centre Ruth Lynden-Bell Scholarship in Scientific Computing. E.K and A.M.E were supported by Ada Lovelace centre at Science and Technology Facilities Council (https://adalovelacecentre.ac.uk/), Physical Sciences Databases Infrastructure (https://psdi.ac.uk, jointly STFC and University of Southampton) under grants EP/X032663/1 and EP/X032701/1,  and EPSRC under grants EP/W026775/1 and EP/V028537/1.

\clearpage

\bibliography{references}

\clearpage

\section{Appendix}

\subsection{Re-estimation of atomic reference energies (E0s)}
\label{sec:e0-est}

During the fine-tuning phase, one needs to adapt the reference energies, to ensure proper normalization of the target energies as the MACE architecture learns to predict relative energies instead of the total energy, as do many other MLIPs. This relative energy is the atomization energy:

\begin{equation}
    E^{\rm atm} = E^{\rm tot}-\sum_i^{N} E^{0}.
\end{equation}

There are currently two choices for E0s: computed isolated atom energies using the same reference method as the training set, or using MACE's ``average'' argument which re-estimates the E0s using averaging.
A linear system can be formulated to provide a more robust approach for E0 re-estimation for fine-tuning. The energy prediction error $\epsilon_i$ for a configuration $i$ is defined as:

\begin{equation}
    \epsilon_i = E^{\text{true}}_i - E^{\text{predicted}}_i.
\end{equation}

\noindent We assume this error can be systematically corrected for each element $j$ by adjusting its value of $E_0$:

\begin{equation}
    \epsilon_i = \sum_{j} N_{ij} \times c_j,
\end{equation}

\noindent where $N_{ij}$ is the number of atoms of element $j$ in configuration $i$ and $c_j$ is the correction for element $j$. In matrix notation, we can write this as $N \mathbf{c} = \boldsymbol{\epsilon}$~\cite{strang2000linear}:

\begin{equation}
\begin{bmatrix}
n_{11} & n_{12} & \cdots & n_{1n} \\
n_{21} & n_{22} & \cdots & n_{2n} \\
\vdots & \vdots & \ddots & \vdots \\
n_{m1} & n_{m2} & \cdots & n_{mn} \\
\end{bmatrix}
\begin{bmatrix}
c_1 \\
c_2 \\
\vdots \\
c_n \\
\end{bmatrix}
=
\begin{bmatrix}
\epsilon_1 \\
\epsilon_2 \\
\vdots \\
\epsilon_m \\
\end{bmatrix}
\end{equation}

Since we are adjusting the E0 values of different elements, which are present in many configurations with different energies, we therefore have an overdetermined system with more equations than unknowns. Since an exact solution may not exist, we can instead minimize the sum of squared residuals, $\min_{\mathbf{x}} ||A\mathbf{x} - \mathbf{b}||^2_2$, and efficiently solve this via the least squares method~\cite{weisberg2005applied}. The least-squares solution follows from the normal equations~\cite{bjorck2024numerical}: 

\begin{equation}
    \mathbf{c} = (N^T N)^{-1} N^T \boldsymbol{\epsilon}.
\end{equation}
Now the E0 values can be re-estimated:
\begin{equation}
    E0_j^{\text{new}} = E0_j^{\text{old}} + c_j.
\end{equation}
Note that the E0s of the replay head are kept fixed to the original DFT values from the pre-training stage.

\section{Hyperparameters}

\begin{table*}[t]
\centering
\caption{Hyper-parameter settings for the three \textsc{MACE} variants used in this work.}
\label{tab:mace_hparams}
\resizebox{\columnwidth}{!}{
\begin{tabular}{@{}lccc@{}}
\toprule
& \multicolumn{3}{c}{\textbf{Models}} \\ 
\cmidrule(lr){2-4}
\textbf{Hyperparameter} & \textbf{mace-omat-1} & \textbf{mace-mh-1} & \textbf{mace-omat-0} \\ 
\midrule
max\_ell              & 3   & 3   & 3   \\
correlation           & 3   & 3   & 3   \\
max\_L                & 2   & 2   & 1   \\
num\_channels\_edge         & 128 & 128 & 128 \\
num\_channels\_node         & 512 & 512 & 128 \\
num\_interactions     & 2   & 2   & 2   \\
num\_radial\_basis    & 8   & 8   & 8   \\
r\_max                & 6   & 6   & 6   \\
interactions\_class   &  non-linear   &  non-linear   &   linear   \\  
irreps                & 16x0e & 16x0e & 16x0e \\
batch size            & 256 & 256 & 256 \\
energy coefficient    & 1 & 1 & 1 \\
force coefficient     & 10  & 10  & 10  \\
stress coefficient & 10 & 10 & 10 \\
\bottomrule
\end{tabular}}
\end{table*}

\paragraph{Training loss}
The models were trained using a weighted sum of Huber losses of energy, forces, and stress:
\begin{equation}
  \begin{aligned}
    \mathcal{L} & = \frac{\lambda_E}{N_b} \sum_{b=1}^{N_b}\mathcal{L}_\text{Huber}\biggl(\frac{\hat{E}_b}{N_a}, \frac{E_b}{N_a}, \delta_E\biggr) \\ 
    &\hphantom{=}+ \frac{\lambda_F}{3\sum_{b=1}^{N_b}N_a} \sum_{b=1}^{N_b} \sum_{a=1}^{N_a}\sum_{i=1}^{3}\mathcal{L}^\star_\text{Huber}\biggl(-\frac{\partial\hat{E}_b}{\partial {r}_{b,a,i}}, F_{b,a,i}, \delta_F\biggr) \\
                & \hphantom{=}+ \frac{\lambda_\sigma}{9N_b} \sum_{b=1}^{N_b}\sum_{i=1}^3\sum_{j=1}^3\mathcal{L}_\text{Huber}\biggl(\frac{1}{V_b}\frac{\partial\hat{E}_b}{\partial {\varepsilon}_{b,ij}}, \sigma_{b,ij}, \delta_\sigma\biggr),
  \end{aligned}
  \label{eq:loss}
\end{equation}
where $\lambda_E, \lambda_F, \lambda_\sigma$ are predetermined weights of energy ($E$), forces ($F$), and stress ($\sigma$) losses, the symbols under a hat correspond to predicted values, and $N_b$ and $N_a$ are the batch size and the number of atoms in each structure. In the last term involving the stress, $\varepsilon_b$ and $\sigma_b$ correspond to the strain and stress tensors, respectively. 
We used $(\lambda_E, \lambda_F, \lambda_\sigma) = (1, 10, 10)$ and Huber deltas of $\delta_E = 0.01, \delta_F = 0.01,  \delta_\sigma = 0.01$.
We use a conditional Huber loss $\mathcal{L}^\star_\text{Huber}$ for forces, where the Huber delta $\delta_F$ is adaptive to the force magnitude on each atom, as used in~\cite{batatia2024foundationmodelatomisticmaterials}.
The Huber delta $\delta_F$ decreases step-wise by a factor from \num{1.0} to \num{0.1} as the atomic force increases from \num{0} to 300 eV/Å.

\paragraph{Optimisation} The models are trained with the AMSGrad~\cite{j.2018on} variant of Adam~\cite{Kingma2014} with default parameters $\beta_1 = 0.9$, $\beta_2 = 0.999$, and $\epsilon=10^{-8}$.

\paragraph{Pre-training}
For the pre-training phase, we use a learning rate of 0.001 and an exponential moving average (EMA) learning scheduler with a decaying factor of 0.999.
We employ gradient clipping of 100.
The model is trained for 7 epochs on 48 NVIDIA H100 GPUs across 12 nodes.

\paragraph{Fine-tuning}

For the fine-tuning phase, we use a learning rate of 0.001 and an exponential moving average (EMA) learning scheduler with a decaying factor of 0.999.
We employ gradient clipping of 100.
The model is trained for 20 epochs on 48 NVIDIA H100 GPUs across 12 nodes.


\begin{table*}[t]
\centering
\caption{Summary of all reported benchmark results for \textbf{mace-mh-0-r2scan} and \textbf{mace-mh-1-r2scan}. When only a ``-D3'' value was reported in the manuscript, it is used directly here (per the instruction to treat D3 and non-D3 as the same).}
\label{tab:mh_r2scan_appendix}
\resizebox{\textwidth}{!}{
\begin{tabular}{@{}llccc@{}}
\toprule
\textbf{Domain} & \textbf{Benchmark / Metric} & \textbf{Unit} & \textbf{mace-mh-0-r2scan} & \textbf{mace-mh-1-r2scan} \\
\midrule
\multirow{6}{*}{Materials} 
& Elastic moduli (no relaxation) – Bulk MAE & GPa & 24.07 & 33.93 \\
& Elastic moduli (no relaxation) – Shear MAE & GPa & 15.59 & 20.80 \\
& Elastic moduli (relaxed) – Bulk MAE & GPa & 12.10 & 20.19 \\
& Elastic moduli (relaxed) – Shear MAE & GPa & 9.03 & 10.60 \\
& Thermal conductivity RMSE & W\,m$^{-1}$K$^{-1}$ & 0.33 & 0.32 \\
& Lattice constants MAE & \AA & 0.048 & 0.027 \\
\midrule
\multirow{7}{*}{Materials (Phonons)}
& $\omega_{\max}$ MAE & K & 21 & 36 \\
& $\omega_{\text{avg}}$ MAE & K & 7 & 10 \\
& $\omega_{\min}$ MAE & K & 14 & 14 \\
& Brillouin zone RMSE & K & 13 & 19 \\
& Entropy $S$ MAE (300 K) & J\,mol$^{-1}$K$^{-1}$ & 16 & 14 \\
& Helmholtz free energy $F$ MAE (300 K) & kJ\,mol$^{-1}$ & 5 & 6 \\
& Heat capacity $C_V$ MAE (300 K) & J\,mol$^{-1}$K$^{-1}$ & 5 & 6 \\
\midrule
\multirow{2}{*}{Molecular Crystals}
& X23 formation energy MAE & kJ\,mol$^{-1}$ & 10.05 & 28.13 \\
& Ice polymorphs relative energy MAE & meV & 7.85 & 25.69 \\
\midrule
\multirow{3}{*}{Surfaces}
& S24 adsorption energy MAE & eV & 0.150 & 0.249 \\
& OC20 adsorption RMSD & eV & 0.242 & 0.185 \\
& OC20 Pearson $r$ & -- & 0.97 & 0.98 \\
\midrule
\multirow{8}{*}{Molecules}
& Wiggle150 strained conformers MAE & kcal\,mol$^{-1}$ & 5.28 & 1.59 \\
& GMTKN55 – Basic (small) MAE & kcal\,mol$^{-1}$ & 16.24 & 12.92 \\
& GMTKN55 – Large systems MAE & kcal\,mol$^{-1}$ & 21.15 & 7.42 \\
& GMTKN55 – Barrier heights MAE & kcal\,mol$^{-1}$ & 11.47 & 8.06 \\
& GMTKN55 – Intramolecular NC MAE & kcal\,mol$^{-1}$ & 14.76 & 10.84 \\
& GMTKN55 – Intermolecular NC MAE & kcal\,mol$^{-1}$ & 13.56 & 12.94 \\
& GMTKN55 – Overall (WTMAD-like) & kcal\,mol$^{-1}$ & 15.17 & 10.72 \\
& S30L MAE & kcal\,mol$^{-1}$ & 10.72 & 10.68 \\
\bottomrule
\end{tabular}
}
\end{table*}


\begin{table*}[h!]
\centering
\caption{MDR Phonon benchmark on the phonon frequencies and thermodynamic properties (300 K) of roughly ten thousand materials. BZ refers to the RMSE across the whole Brillouin Zone.}
\label{tab:mdr_benchmark_full}
\resizebox{\columnwidth}{!}{%
\begin{tabular}{@{}lccccccc@{}}
\toprule
\textbf{Material} & $\omega_{\text{max}}$ & $\omega_{\text{avg}}$ & $\omega_{\text{min}}$ & BZ & $S$ & $F$ & $C_V$ \\
& (K) & (K) & (K) & (K) & (J/mol$\cdot$K) & (kJ/mol) & (J/mol$\cdot$K) \\
\midrule
uma-s-1p1-omat    & 25  & \textbf{3} & 138 & 16 & 13 & 3 & 6  \\
uma-m-1p1-omat    & 17  & \textbf{3} & 101 & 13 & 11 & \textbf{2} & 5  \\
orb-v3-consv-inf-omat   & \textbf{10}  & \textbf{3} & 38 & 10 & 9 & \textbf{2} & 4  \\
mace-omat-0     & 16 & 4 & 13 & 10 & 10 & 3 & 3 \\
mace-mp-0a     & 65  & 32 & 19 & 41 & 60 & 23 & 14 \\
mattersim-5M   & 19  & 5 & 16 & 13 & 14 & 4 & 4  \\
mace-mh-1-omat  & 12  & \textbf{3} & \textbf{11} & \textbf{7} & 8 & \textbf{2} & 3  \\
mace-omat-1 & 13  & \textbf{3} & 12 & 8 & 8 & \textbf{2} & 3 \\
uma-s-1p1-omat (no-sym)    & 11  & 4 & 21 & 14 & \textbf{7} & \textbf{2} & 3  \\
orb-v3-consv-inf-omat (no-sym)   & 12  & 5 & 29 & 24 & 13 & 3 & 4  \\
uma-m-1p1-omat (no-sym)    & \textbf{9}  & \textbf{3} & 18 & 13 & 8 & \textbf{2} & \textbf{2}  \\
mattersim-5M (no-sym) & 20 & 6 & 18 & 16 & 13 & 4 & 3 \\
mace-omat-0  (no-sym) & 18 & 5 & 16 & 13 & 10 & 3 & 3 \\
mace-mp-0a (no-sym) & 67 & 33 & 21 & 43 & 60 & 24 & 13 \\
mace-mh-1-omat  (no-sym) & 13 & 4 & 13 & 10 & 8 & 3 & \textbf{2} \\
mace-omat-1 (no-sym) & 15 & 4 & 13 & 11 & 8 & 3 & \textbf{2} \\
\bottomrule
\end{tabular}%
}
\end{table*}
\clearpage

\begin{table*}[t]
\centering
\setlength{\tabcolsep}{4pt}
\caption{Normalization bounds for all benchmarks. Direction: Lower-is-better (L) or Higher-is-better (H).}
\label{tab:normalization_bounds}
\resizebox{\textwidth}{!}{%
\begin{tabular}{@{}llllcc l@{}}
\toprule
\textbf{Group} & \textbf{Benchmark / Metric} & \textbf{Unit} & \textbf{Dir.} & \textbf{Best (b)} & \textbf{Worst (w)} & \textbf{Rationale} \\
\midrule
Materials & Elastic moduli (bulk) MAE & GPa & L & 0.0 & 50.0 & Mathematical (zero error) \\
Materials & Elastic moduli (shear) MAE & GPa & L & 0.0 & 50.0 & Mathematical (zero error) \\
Materials & Thermal conductivity RMSE & W\,m$^{-1}$K$^{-1}$ & L & 0.0 & 2.0 & Mathematical (zero error) \\
Materials & Phonon $\omega_{\max}$ MAE & K & L & 0.0 & 50.0 & Mathematical (zero error) \\
Materials & Phonon $\omega_{\text{avg}}$ MAE & K & L & 0.0 & 50.0 & Mathematical (zero error) \\
Materials & Phonon $\omega_{\min}$ MAE & K & L & 0.0 & 50.0 & Mathematical (zero error) \\
Materials & Phonon BZ MAE & K & L & 0.0 & 50.0 & Mathematical (zero error) \\
Materials & Entropy MAE & J\,mol$^{-1}$K$^{-1}$ & L & 0.0 & 50.0 & Mathematical (zero error) \\
Materials & Helmholtz free energy MAE & kJ\,mol$^{-1}$ & L & 0.0 & 50.0 & Mathematical (zero error) \\
Materials & Heat capacity $C_V$ MAE & J\,mol$^{-1}$K$^{-1}$ & L & 0.0 & 50.0 & Mathematical (zero error) \\
Materials & Phonon stability & count & L & 0.0 & 50.0 & Mathematical (zero instabilities) \\
Materials & Phonon stability $n$ & count & L & 0.0 & 50.0 & Mathematical (zero instabilities) \\
Molecular Crystals & X23 formation energy MAE & kcal\,mol$^{-1}$ & L & 0.5 & 50.0 & Reference accuracy (0.5 kcal\,mol$^{-1}$) \\
Molecular Crystals & Ice polymorphs energy MAE & kcal\,mol$^{-1}$ & L & 0.5 & 50.0 & Reference accuracy (0.5 kcal\,mol$^{-1}$) \\
Surfaces & S24 adsorption energy MAE & kcal\,mol$^{-1}$ & L & 0.5 & 50.0 & Reference accuracy (0.5 kcal\,mol$^{-1}$) \\
Surfaces & OC20 adsorption MAE & eV & L & 0.021 & 0.50 & Reference accuracy (0.5 kcal\,mol$^{-1}$) \\
Molecules & Wiggle150 MAE & kcal\,mol$^{-1}$ & L & 0.5 & 50.0 & Reference accuracy (0.5 kcal\,mol$^{-1}$) \\
Molecules & GMTKN55 Basic small MAE & kcal\,mol$^{-1}$ & L & 0.5 & 50.0 & Reference accuracy (0.5 kcal\,mol$^{-1}$) \\
Molecules & GMTKN55 Large systems MAE & kcal\,mol$^{-1}$ & L & 0.5 & 50.0 & Reference accuracy (0.5 kcal\,mol$^{-1}$) \\
Molecules & GMTKN55 Barrier heights MAE & kcal\,mol$^{-1}$ & L & 0.5 & 50.0 & Reference accuracy (0.5 kcal\,mol$^{-1}$) \\
Molecules & GMTKN55 Intra-NC MAE & kcal\,mol$^{-1}$ & L & 0.5 & 50.0 & Reference accuracy (0.5 kcal\,mol$^{-1}$) \\
Molecules & GMTKN55 Inter-NC MAE & kcal\,mol$^{-1}$ & L & 0.5 & 50.0 & Reference accuracy (0.5 kcal\,mol$^{-1}$) \\
Molecules & GMTKN55 overall MAE & kcal\,mol$^{-1}$ & L & 0.5 & 50.0 & Reference accuracy (0.5 kcal\,mol$^{-1}$) \\
Molecules & PLF547 MAE & kcal\,mol$^{-1}$ & L & 0.5 & 50.0 & Reference accuracy (0.5 kcal\,mol$^{-1}$) \\
Molecules & S30L MAE & kcal\,mol$^{-1}$ & L & 0.5 & 50.0 & Reference accuracy (0.5 kcal\,mol$^{-1}$) \\
Physicality & Slab extensivity $\Delta$ & meV & L & 0.0 & 50.0 & Mathematical (extensivity; zero deviation) \\
Physicality & H-additivity max $|\Delta F|$ & meV\,\AA$^{-1}$ & L & 0.0 & 50.0 & Mathematical (additivity; zero deviation) \\
Physicality & Diatomic force flips & count & L & 1.0 & 10.0 & Mathematical/physical (one sign change) \\
Physicality & Diatomic energy minima & count & L & 1.0 & 10.0 & Mathematical/physical (single minimum) \\
Physicality & Diatomic inflections & count & L & 2.0 & 10.0 & Mathematical/physical (typical Morse shape) \\
Physicality & Spearman $\rho(E_{\rm rep})$ & -- & H & $-1.0$ & $-0.5$ & Mathematical (perfect anticorrelation) \\
Physicality & Spearman $\rho(E_{\rm at})$ & -- & H & 1.0 & 0.0 & Mathematical (perfect correlation) \\
\bottomrule
\end{tabular}
}
\end{table*}

\end{document}